\documentclass[11pt]{article}
\usepackage{amsmath}
\usepackage{graphicx}
\usepackage{epsfig}

\def\simgt{\mathrel{\lower2.5pt\vbox{\lineskip=0pt\baselineskip=0pt
           \hbox{$>$}\hbox{$\sim$}}}}
\def\simlt{\mathrel{\lower2.5pt\vbox{\lineskip=0pt\baselineskip=0pt
           \hbox{$<$}\hbox{$\sim$}}}}
\newcommand{\notEt}{E_{T}\kern-1.21em\hbox{/}\kern0.45em}

\begin{document}
\baselineskip 0.5cm

\begin{titlepage}

\begin{flushright}
\end{flushright}

\vskip 2.0cm

\begin{center}

{\Large \bf 
Environmentally Selected WIMP Dark Matter \\ \vspace{-0.08cm}
with \\
High-Scale Supersymmetry Breaking}

\vskip 0.6cm

{\large Gilly Elor$^a$, Hock-Seng Goh$^a$, Lawrence J. Hall$^{a,b}$, 
Piyush Kumar$^a$, and Yasunori Nomura$^{a,b}$}

\vskip 0.4cm

$^a$ {\it Berkeley Center for Theoretical Physics, \\
     and Theoretical Physics Group, Lawrence Berkeley National Laboratory \\
     University of California, Berkeley, CA 94720, USA} \\
$^b$ {\it Institute for the Physics and Mathematics of the Universe, \\
     University of Tokyo, Kashiwa 277-8568, Japan}

\begin{abstract}
We explore the possibility that both the weak scale and the thermal 
relic dark matter abundance are environmentally selected in a multiverse. 
An underlying supersymmetric theory containing the states of the MSSM 
and singlets, with supersymmetry and $R$ symmetry broken at unified 
scales, has just two realistic low energy effective theories.  One 
theory, (SM~$+~\tilde{w}$), is the Standard Model augmented only by 
the wino, having a mass near $3~{\rm TeV}$, and has a Higgs boson 
mass in the range of $(127~\mbox{--}~142)~{\rm GeV}$.  The other theory, 
(SM~$+~\tilde{h}/\tilde{s}$), has Higgsinos and a singlino added to 
the Standard Model.  The Higgs boson mass depends on the single new 
Yukawa coupling of the theory, $y$, and is near $141~{\rm GeV}$ for 
small $y$ but grows to be as large as $210~{\rm GeV}$ as this new 
coupling approaches strong coupling at high energies.  Much of the 
parameter space of this theory will be probed by direct detection 
searches for dark matter that push two orders of magnitude below the 
present bounds; furthermore, the dark matter mass and cross section 
on nucleons are correlated with the Higgs boson mass.  The indirect 
detection signal of monochromatic photons from the galactic center is 
computed, and the range of parameters that may be accessible to LHC 
searches for trilepton events is explored.  Taking a broader view, 
allowing the possibility of $R$ symmetry protection to the TeV scale 
or axion dark matter, we find four more theories: (SM~+~axion), 
two versions of Split Supersymmetry, and the E-MSSM, where a little 
supersymmetric hierarchy is predicted.  The special Higgs mass 
value of $(141 \pm 2)~{\rm GeV}$ appears in symmetry limits 
of three of the six theories, (SM~+~axion), (SM~$+~\tilde{w}$) 
and (SM~$+~\tilde{h}/\tilde{s})$, motivating a comparison of other 
signals of these three theories.
\end{abstract}

\end{center}
\end{titlepage}

\newpage

\vspace{-1.0cm} 
\tableofcontents

\section{Introduction and Framework}
\label{sec:intro}

The extremely small value of the cosmological constant---at least $120$ 
orders below its natural value---has resisted explanation from symmetry 
principles.  Furthermore, since the discovery of dark energy, with 
$w = -1.0 \pm 0.1$~\cite{Perlmutter:1998np,Komatsu:2008hk}, there 
appears to be a second part to the cosmological constant problem: why 
does it take the observed value, which is close to the matter density 
at the present epoch?  Environmental arguments solve both pieces of the 
puzzle, at least at the order of magnitude level~\cite{Weinberg:1987dv}, 
motivating searches for further evidence of environmental selection. 
While there are symmetry arguments for understanding why the weak 
scale is $16$ orders of magnitude less than its natural value in the 
Standard Model (SM), no direct experimental evidence for any such 
symmetry extension of the SM has been found so far in the preliminary 
exploration of the weak scale.  On the other hand, it is intriguing 
that if the weak scale were increased by a factor of $2$, the SM would 
not yield any complex stable nuclei~\cite{Agrawal:1997gf}, suggesting 
that the weak scale may also be selected environmentally.

Such arguments from environmental selection only make sense if the 
underlying theory of nature has a vast landscape of vacua, allowing 
a fine scanning of parameters.  The realization that string theory, 
with a compact manifold of extra dimensions, may yield such a 
landscape~\cite{Bousso:2000xa} strengthens the motivation for 
seeking further evidence for environmental selection on a multiverse. 
String theory requires supersymmetry; however, if the weak scale 
is environmental, it is reasonable to consider that the scale of 
supersymmetry breaking is high, i.e.\ close to the string scale rather 
than to the weak scale.  Hence, in this paper we explore the case of 
very high scale supersymmetry breaking, with the physics of electroweak 
symmetry breaking described purely by the SM Higgs boson.

This appears to leave no room for weakly interacting massive particle 
(WIMP) dark matter, since no new stable particles have masses that 
are logically connected to weak symmetry breaking.  Instead, we 
{\it assume} that there is an environmental requirement for dark 
matter.  This assumption now plays a central role, since it determines 
the properties of the dark matter and the size of the cosmological 
constant.  In particular, we assume that dark matter arises from 
a conventional freezeout process through interactions with couplings 
that are order unity.  The environmentally allowed range of the dark 
matter density is unknown~\cite{Tegmark:2005dy}, but we do not need to 
assume that it is very narrow.  Even if the range spans several orders 
of magnitude, the selection process leads to dark matter particles 
having masses within an order of magnitude or two of the weak scale. 
However, we stress that the connection between the dark matter mass, 
$m_{\rm DM}$, and the weak scale, $v$, is now purely coincidental.

Our framework for high scale supersymmetry breaking, with independent 
environmental selections of the weak scale and dark matter, is as follows. 
The scale of supersymmetry breaking in the SM sector, $\tilde{m}$, is 
taken close to the cutoff of the effective field theory at the string 
scale, $M_*$.  At $\tilde{m}$ the theory is taken to have the states 
of the Minimal Supersymmetric Standard Model (MSSM) together with gauge 
singlet chiral multiplets $S$.  The environmental selection for dark 
matter requires some states beyond those of the SM to be much lighter 
than $\tilde{m}$. As supersymmetry breaks near $M_*$, the $R$ symmetry 
$U(1)_R$ must also be broken at a very high scale to cancel the cosmological 
constant.  It is then likely that $R$ breaking is transmitted to the 
MSSM sector without large suppression, making the gauginos very heavy. 
It is possible that such a strong transmission is avoided, leading to 
the Split Supersymmetry theory~\cite{ArkaniHamed:2004fb}.  In this case, 
environmental selection of dark matter acting on the symmetry breaking 
parameter of a particular $R$ symmetry can protect all fermionic 
superpartners to near the weak scale.

In this paper we explore the simple possibility that $R$ breaking in 
the MSSM sector is unsuppressed.  In this case, what is the origin of 
the small weak-scale mass needed for WIMP dark matter?  Environmental 
selection could lead either to a cancellation of terms in the mass of 
some particle, or to a theory with an approximate symmetry that protects 
some light states.  In the former case dark matter would be fermionic, 
since for fermions the cancellation is linear in the mass, while for 
scalars it is quadratic.  Also, the dark matter sector would contain 
just a single fermion, since there is a cost for each cancellation. 
Since pure Higgsino dark matter is observationally excluded and 
a pure singlino does not freeze-out at the weak scale, in our framework 
the only possibility with mass cancellation is pure wino dark matter, 
$\tilde{w}$.  If the mass of the dark matter sector is protected by 
an approximate symmetry, there is again a unique possibility.  Since 
supersymmetry is broken at the high scale, the approximate symmetry 
cannot keep scalars light.  Furthermore, since the approximate symmetry 
must be a non-$R$ symmetry it cannot protect gauginos either, so that 
the only states that it can protect are the Higgsinos and singlinos, 
$\tilde{h}_u$, $\tilde{h}_d$ and $\tilde{s}$.  Realistic WIMP dark 
matter requires that all three states are involved.

Thus with high scale breaking of the $R$ symmetry, we are led to two 
possibilities for environmental WIMP dark matter: selection of a small 
wino mass, or selection of an approximate symmetry that protects the 
Higgsinos and singlino.  We do not know which of these possibilities 
has a higher probability in the multiverse.
\begin{figure}[t]
  \center{\includegraphics[scale=0.9]{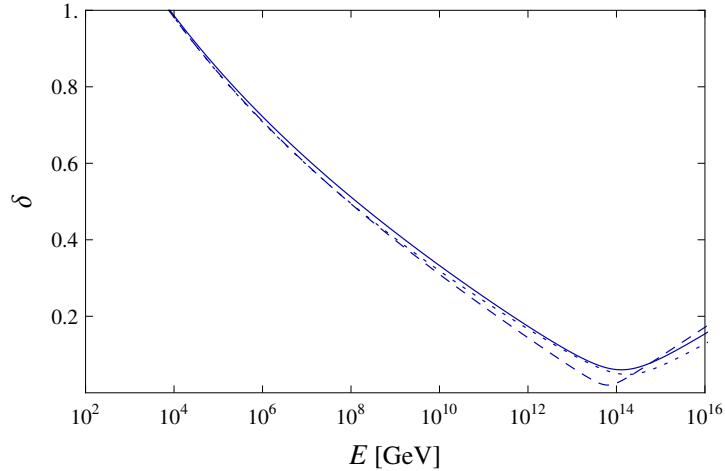}}
\caption{The threshold correction $\delta$ required for gauge coupling 
 unification at energy $E$ in the SM (solid), in SM~$+~\tilde{h}/\tilde{s}$ 
 (dashed), and in SM~$+~\tilde{w}$ (dotted).  All three theories unify 
 near $10^{14}~{\rm GeV}$, with varying levels of precision.  Here 
 $\delta \equiv \sqrt{(g_1^2-\bar{g}^2)^2 + (g_2^2-\bar{g}^2)^2 
 + (g_3^2-\bar{g}^2)^2}/\bar{g}^2$, where $\bar{g}^2 \equiv 
 (g_1^2+g_2^2+g_3^2)/3$.}
\label{fig:delta}
\end{figure}
In Figure~\ref{fig:delta}, we illustrate how accurately gauge coupling 
unification occurs in these two theories compared to the SM.  The 
quantity $\delta(E)$ is a measure of the size of the threshold 
corrections needed to unify the couplings at scale $E$.  In all 
three theories unification occurs near $10^{14}~{\rm GeV}$.  The SM 
and the SM~$+~\tilde{w}$ theory require comparable threshold corrections 
of $(5~\mbox{--}~6)\%$, while SM~$+~\tilde{h}/\tilde{s}$ yields 
gauge coupling unification that is almost as precise as in the 
MSSM~\cite{ArkaniHamed:2005yv,Mahbubani:2005pt}.  Since gauge 
coupling unification occurs at a scale $M_u \sim 10^{14}~{\rm GeV}$, 
suppression of gauge-mediated proton decay suggests unification into 
higher dimensions~\cite{Hall:2001pg}, which can bring gauge coupling 
unification for the SM~$+~\tilde{h}/\tilde{s}$ theory to an even 
higher level of precision~\cite{Hall:2001xb,Mahbubani:2005pt}. 
We study both these theories in this paper; various aspects of the 
SM~$+~\tilde{h}/\tilde{s}$ theory and its signals are explored in 
sections~\ref{sec:model}~--~\ref{sec:signals}, and wino dark matter 
in the SM~$+~\tilde{w}$ theory is examined in section~\ref{sec:wino}.

In previous studies of mixed doublet/singlet fermion dark matter inspired by 
an environmental weak scale, the theory studied was the most general allowed 
for the doublet/singlet system~\cite{ArkaniHamed:2005yv,Mahbubani:2005pt}. 
The Higgsino/singlino dark matter theory studied in this paper is a new 
version, which emerges from the supersymmetric framework described above 
and keeps the successful features of gauge coupling unification and 
WIMP dark matter.  Supersymmetry plays a key role since it leads 
to a high precision prediction for the Higgs boson mass, and it 
restricts the form of the Yukawa couplings to the new states.  In this 
SM~$+~\tilde{h}/\tilde{s}$ theory, the Higgsino mass, $\mu$, transforms 
non-trivially under some approximate symmetry $G$ and, since $G$ is not 
an $R$ symmetry, the Higgs-boson mass mixing parameter, $B_\mu$, must 
also be suppressed: $\mu \sim \epsilon \tilde{m}$ and $B_\mu \sim 
\epsilon \tilde{m}^2$, where $\epsilon$ is a small symmetry breaking 
parameter.  This implies that environmental selection on the Higgs mass 
matrix to obtain a low weak scale leads to the light Higgs boson being 
$h \sim h_u + \epsilon h_d$ or $h \sim h_d + \epsilon h_u$, with 
$\epsilon \sim m_{\rm DM}/\tilde{m} \sim O(10^{-12}~\mbox{--}~10^{-11})$. 
The heavy top quark prefers the former case, so that quark and charged 
lepton masses arise from
\begin{equation}
  {\cal L} \sim [QUH_u]_{\theta^2} 
    + \frac{1}{M_*^2} [(QD + LE) H_u^\dagger X^\dagger]_{\theta^4} 
    + {\rm h.c.},
\label{eq:SM-Yukawa}
\end{equation}
where $Q,U,D,L,E$ are the quark and lepton superfields, $H_u$ a Higgs 
superfield and $X$ a superfield that leads to supersymmetry breaking, 
$\langle X \rangle = \theta^2 F_X$ with $F_X \sim \tilde{m} M_*$.  Hence 
the $b$ quark and $\tau$ lepton masses are moderately suppressed relative 
to the $t$ quark mass by supersymmetry breaking, $m_{b,\tau}/m_t \sim 
\tilde{m}/M_*$.

As in the MSSM, supersymmetry imposes a boundary condition on the 
Higgs quartic coupling, but with two important differences.  First, 
the boundary condition applies near the unified scale rather than near 
the weak scale and, as we will see in section~\ref{sec:higgs}, this 
has crucial consequences for the prediction of the Higgs boson mass. 
Secondly, since $\tan\beta \sim 1/\epsilon$, the boundary condition 
becomes independent of $\tan\beta$
\begin{equation}
  \lambda(\tilde{m}) = \frac{g^2(\tilde{m}) + g'^2(\tilde{m})}{8}\, 
    (1 + \delta),
\label{eq:lambdabc}
\end{equation}
where $\delta$ results from threshold corrections, for example, from 
integrating out superpartners such as the top squarks.

Since $\tilde{s}$ is lighter than $\tilde{m}$, the superfield $S$ 
must also transform non-trivially under $G$.  This removes interactions 
that transform linearly in $S$, such as $[S X^\dagger X]_{\theta^4}$. 
The $G$ charges must allow $[S H_u H_d]_{\theta^2}$, since otherwise 
$\tilde{s}$ would not interact in the low energy theory, preventing 
acceptable dark matter.  Hence the theory below $\tilde{m}$ is 
described by
\begin{equation}
  {\cal L} = {\cal L}_{\rm SM}(q,u,d,l,e,h) 
    + \Bigl\{ \mu \tilde{h}_u \tilde{h}_d + \frac{m}{2} \tilde{s}^2 
      + y \tilde{h}_d \tilde{s} h + {\rm h.c.} \Bigr\},
\label{eq:L}
\end{equation}
where we drop higher dimension interactions and dimensionless interactions 
suppressed by powers of $\epsilon$.  Supersymmetry together with $G$ 
removes the gauge invariant interaction $\tilde{h}_u \tilde{s} h^\dagger$, 
so that there are just three new parameters: $\mu, m$ and $y$.  Since 
the quartic coupling is predicted in Eq.~(\ref{eq:lambdabc}), the theory 
has just two more free parameters than the SM.

In the next section we present a particular realization of 
our SM~$+~\tilde{h}/\tilde{s}$ theory, with $G = Z_3$.  In 
section~\ref{sec:higgs} we present the Higgs mass prediction, finding 
that it is extremely insensitive to $y$ for $y(\tilde{m}) \simlt 0.4$. 
The constraints on the parameter space imposed by the requirement of dark 
matter are studied in section~\ref{sec:DM}.  In section~\ref{sec:signals} 
we study experimental signals from hadron colliders and from the direct 
and indirect detection of dark matter.  In section~\ref{sec:wino} we 
study wino dark matter in the SM~$+~\tilde{w}$ theory.  Discussion and 
conclusions are given in section~\ref{sec:concl}.

\section{A Model with Comparable Higgsino and Singlino Masses}
\label{sec:model}

In this section we present a model that realizes the framework where dark 
matter is protected by a chiral symmetry $G$.  We consider that the theory 
above $\tilde{m}$ is the MSSM with a singlet superfield $S$.  We also 
assume that the theory possesses usual $R$ parity, under which $S$ is 
even, and that supersymmetry breaking does not violate $G$ so that the 
protection of dark matter persists below $\tilde{m}$.  This implies 
that the supersymmetry breaking field $X$ is neutral under $G$.

We consider that the Higgsinos and singlino obtain masses in the same 
way via a single symmetry breaking spurion $\epsilon$: ${\cal L} \sim 
[\epsilon M_* H_u H_d + \epsilon M_* S^2]_{\theta^2}$.  We also require 
that the Yukawa coupling $[S H_u H_d]_{\theta^2}$ exists, allowing the 
Higgsinos to mix with the singlino.  We then find, using the freedom 
of redefining $G$ charges through hypercharge, that $G = Z_3$ is the 
unique possibility under which $H_u$, $H_d$, and $S$, as well as the 
symmetry breaking spurion $\epsilon$, all carry a charge of $+1$. 
The existence of the Yukawa couplings of Eq.~(\ref{eq:SM-Yukawa}) 
then implies that the charge of $QU$, $QD$, and $LE$ are $-1$, $+1$, 
and $+1$, respectively.  The most general charge assignment under 
$G$ is thus given as in Table~\ref{tab:Z_3}. 
\begin{table}[t]
\begin{center}
\begin{tabular}{|l||ccccc|ccc|c|}
\hline
 & $Q$ & $U$ & $D$ & $L$ & $E$ & $H_u$ & $H_d$ & $S$ & $\epsilon$ \\
\hline
 $Z_3$ & $q$ & $-1-q$ & $1-q$ & $l$ & $1-l$ & $1$ & $1$ & $1$ & $1$ \\
\hline
\end{tabular}
\end{center}
\caption{The charge assignment under $G = Z_3$, where $q$ and $l$ are 
 arbitrary numbers.}
\label{tab:Z_3}
\end{table}

Because of supersymmetry breaking, all the gaugino and scalar fields 
decouple at the scale $\tilde{m}$, except for a single Higgs doublet 
required by environmental selection.  All the supersymmetry breaking 
parameters are of order $\tilde{m}$, except for the holomorphic squared 
masses for $H_{u,d}$ and $S$ (and a linear term for $S$; see below), 
which are suppressed by $\epsilon$.  This implies that the Higgs 
mass-squared matrix takes the form
\begin{equation}
  {\cal L} \sim   \left( \begin{array}{cc}
    h_u^\dagger & h_d 
  \end{array} \right) 
  \left( \begin{array}{cc}
    \tilde{m}_2^2 & \epsilon \tilde{m}_3^2 \\
    \epsilon \tilde{m}_3^2 & \tilde{m}_1^2 
  \end{array} \right) 
  \left( \begin{array}{c}
    h_u \\ h_d^\dagger 
  \end{array} \right),
\label{eq:H-matrix}
\end{equation}
where $\tilde{m}_{1,2,3}^2$ are typically of order $\tilde{m}^2$ and 
scan independently in the multiverse.  Given that environmental selection 
requires one eigenvalue of this matrix to be of order $v^2$, what is 
the most probable linear combination left at low energies?  We find 
that the low energy doublet, $h$, is mostly $h_u$ (or $h_d$) with only 
a tiny, $O(\epsilon)$ mixture of $h_d$ ($h_u$).  Motivated by the large 
top quark mass, we consider
\begin{equation}
  h \sim h_u + \epsilon h_d,
\label{eq:LE-h}
\end{equation}
which corresponds to the case where the environmental requirements selects 
$\tilde{m}_2^2 \sim O(\epsilon^2 \tilde{m}^2)$ and $\tilde{m}_1^2 \sim 
\tilde{m}_3^2 \sim O(\tilde{m}^2)$.  This implies that the low-energy 
SM Higgs doublet is essentially $h_u$ in this model.

The charge assignment of Table~\ref{tab:Z_3} allows operators such as 
$[\epsilon^2 M_*^2 S]_{\theta^2}$, $[S^3]_{\theta^2}$, and $[\epsilon 
X^\dagger X S^\dagger/M_*]_{\theta^4}$.  In particular, the last operator 
induces a vacuum expectation value of the $S$ field, of order $\langle S 
\rangle \sim \epsilon M_*$.  This is, however, not a problem, and the 
low-energy Lagrangian still takes the form of Eq.~(\ref{eq:L}), with $\mu 
\sim m \sim O(\epsilon M_*)$.  The environmental requirement for dark matter 
will then select $\epsilon M_* \sim O(100~{\rm GeV}~\mbox{--}~{\rm TeV})$. 
The Yukawa coupling $y$ is not suppressed by $\epsilon$ and is generically 
expected to be of order unity.

The other possible Yukawa coupling $\tilde{h}_u \tilde{s} h^\dagger$, 
allowed by gauge invariance, is suppressed by $\epsilon$.  This has an 
important consequence that the new fermionic sector at the weak scale, 
$\tilde{h}_{u,d}$ and $\tilde{s}$, does not give any appreciable electric 
dipole moments for the SM particles, since all the phases of $\mu$, $m$ 
and $y$ can be absorbed by field redefinitions of $\tilde{h}_{u,d}$ and 
$\tilde{s}$.  An observation of nonzero electric dipole moments for the 
electron, neutron or atoms in future experiments would therefore exclude 
the model.

Neutrino masses can be obtained by choosing $l = -1$ (mod~3) through 
operators $[(L H_u)^2/M_*]_{\theta^2}$.  Since $M_*$ is expected not far 
from the scale of gauge coupling unification, $\approx 10^{14}~{\rm GeV}$, 
this gives the desired size for the masses.  Alternatively, we can introduce 
right-handed neutrino superfields with the charge $N(-1-l)$ and the Yukawa 
couplings $[L N H_u]_{\theta^2}$.  For $l = -1$ (mod~3), this allows 
the conventional seesaw mechanism.  For $l+1 \neq 0, \pm 1/2, \pm 1, 3/2$, 
the Majorana masses for $N$ are forbidden; but Dirac neutrino masses of 
the right size can still be obtained if we take the Yukawa couplings to 
be of $O(10^{-13}~\mbox{--}~10^{-12})$.

The choice for $q$ is arbitrary.  For $3q+l \neq 0$ (mod 3), proton decay 
is (almost) completely suppressed.  This does not contradict gauge coupling 
unification, if grand unification is realized in higher dimensions and 
matter fields propagate in the bulk~\cite{Hall:2001pg}.

\section{The Higgs Boson Mass}
\label{sec:higgs}

The supersymmetric boundary condition of Eq.~(\ref{eq:lambdabc}) on 
the Higgs quartic coupling $\lambda$, leads to a Higgs mass prediction 
in the SM~$+~\tilde{h}/\tilde{s}$ theory described by Eq.~(\ref{eq:L}). 
In this section we compute the Higgs boson mass, paying attention to 
a variety of uncertainties.

Gauge coupling unification occurs with high precision in the 
SM~$+~\tilde{h}/\tilde{s}$ theory, determining the scale of unification 
$M_u$ to be within an order of magnitude of $10^{14}~{\rm GeV}$.  This 
scale may be closely related to a compactification scale of the compact 
manifold of string theory, and may not be far below the field theory 
cutoff scale $M_*$.  Since we are taking the supersymmetry breaking 
scale $\tilde{m}$ to be very high, could it be close to $M_*$ in 
such a way that the supersymmetric boundary condition on $\lambda$ 
is destroyed?  Naively the supersymmetry breaking corrections to the 
boundary condition are $\delta\lambda \sim (\tilde{m}/M_*)^2$.  In fact, 
in many theories the corrections are suppressed, even as $\tilde{m}$ 
approaches $M_*$~\cite{Hall:2009nd}.  This is because the large value 
of the Planck scale, $M_{\rm Pl} \gg M_*$, implies a large volume 
for the compact manifold.  For example, suppose that supersymmetry 
breaking is localized in the bulk and separated from the location of 
the Higgs multiplet.  In this case, supersymmetry breaking is transmitted 
non-locally; gravitational mediation is suppressed by the volume of 
the bulk, while gauge mediations is loop suppressed.  Alternatively, 
supersymmetry breaking may occur through a nontrivial boundary condition 
in the manifold.  The corrections to the Higgs quartic coupling are 
also volume suppressed in this case.

\begin{figure}[t]
  \center{\includegraphics[scale=1.15]{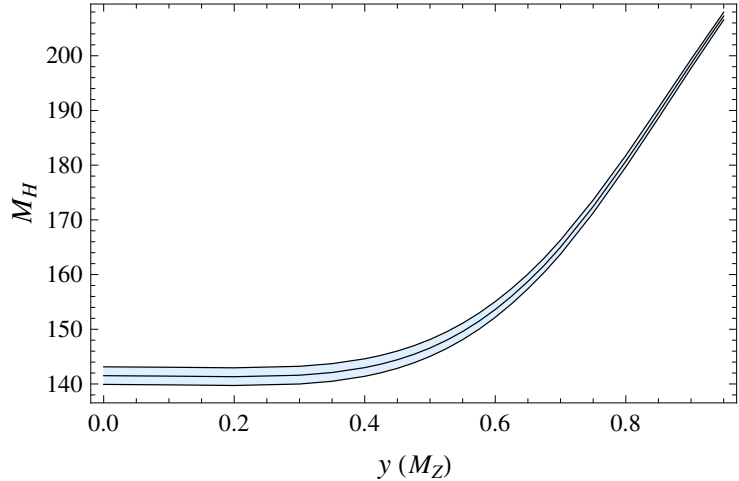}}
\caption{The Higgs mass prediction in the SM~$+~\tilde{h}/\tilde{s}$ 
 theory as a function of $y(M_Z)$.  The curve represents the prediction 
 for $m_t = 173.1~{\rm GeV}$, with the shaded band corresponding to 
 the uncertainty that arises from the experimental error in the top 
 quark mass of $\pm 1.3~{\rm GeV}$.  The supersymmetry breaking scale 
 and QCD coupling are fixed at $\tilde{m} = 10^{14}~{\rm GeV}$ and 
 $\alpha_s(M_Z) = 0.1176$, respectively.}
\label{fig:M_H-1}
\end{figure}
Our basic result for the Higgs boson mass is shown in Figure~\ref{fig:M_H-1} 
where, motivated by the scale of gauge coupling unification, we have 
taken $\tilde{m} = 10^{14}~{\rm GeV}$.  We have assumed that states 
associated with unification are above $\tilde{m}$, so that the theory 
below $\tilde{m}$ is the SM~$+~\tilde{h}/\tilde{s}$ theory with the single 
unknown parameter $y$.  The Higgs mass prediction is shown for the full 
range of $y(M_Z)$ that maintains $y$ perturbative to $\tilde{m}$.  All 
figures and analytical results are obtained using two-loop renormalization 
group (RG) scaling of all couplings from $\tilde{m}$ to the weak scale, 
together with one-loop threshold corrections at the weak scale, including 
the one-loop effective potential for the Higgs field.  In addition, we 
include the two- and three-loop QCD threshold corrections in converting 
the top-quark pole mass to the $\overline{\rm MS}$ top Yukawa coupling, 
since they are anomalously large.

The prediction of Figure~\ref{fig:M_H-1} is shown as a shaded band 
that has a width corresponding to the present experimental error in the 
top quark mass, $m_t = (173.1 \pm 1.3)~{\rm GeV}$~\cite{Tevatron:2009ec}. 
The QCD coupling is fixed at $\alpha_s(M_Z) = 0.1176$~\cite{Amsler:2008zzb}; 
the uncertainty in the Higgs boson mass arising from the experimental 
error, $\pm 0.002$,  of $\alpha_s(M_Z)$ is smaller than that from the 
$m_t$ error by about a factor of $2$.  In the ``small $y$'' region, 
$y \simlt 0.4$, the prediction is almost independent of $y$: $M_H 
\simeq (141~\mbox{--}~142)~{\rm GeV}$, with an error of about $1.5\%$ 
arising from the $m_t$ and $\alpha_s$ uncertainties.  In this region 
the prediction is very close to the case that only the SM survives 
below $\tilde{m}$~\cite{Hall:2009nd}; the coupling $y$ has a negligible 
effect and the Higgsino contributions to the gauge beta functions 
raise the Higgs mass by only about half a GeV.  In the ``large $y$'' 
region, $y \simgt 0.7$, the Higgs mass prediction rises linearly 
with $y(M_Z)$, reaching a maximum value of $\approx 210~{\rm GeV}$. 
Much of the RG scaling of the quartic coupling now arises from $y$, 
reducing the sensitivity to the top Yukawa coupling, as shown by 
the narrowing of the shaded band as $y$ increases.

Two important sources of corrections to the prediction arise from 
threshold corrections to the supersymmetric boundary condition, 
$\delta$ in Eq.~(\ref{eq:lambdabc}), and the choice of matching 
scale, $\tilde{m}$.  In fact, one needs a prescription to determine 
how much of the threshold corrections are absorbed into the matching 
scale.  We absorb all leading-log threshold corrections into the 
matching scale, so that $\delta$ contains only the non-log terms.

\begin{figure}[t]
  \center{\includegraphics[scale=1.15]{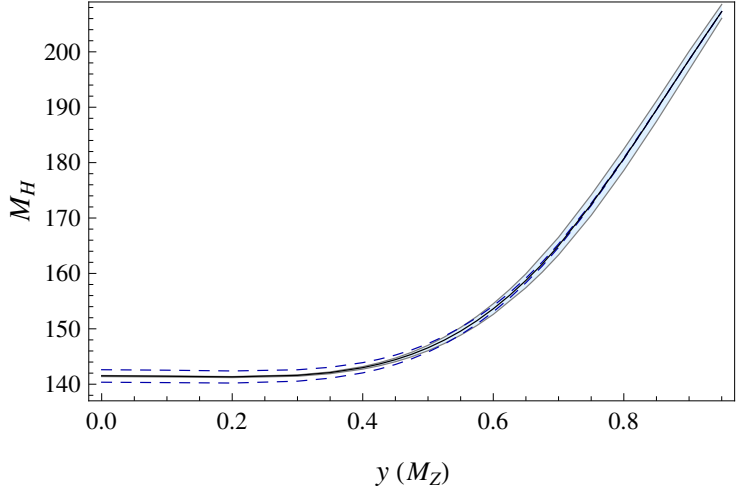}}
\caption{The sensitivity of the Higgs mass prediction to $\tilde{m}$ 
 and $\delta$ in the SM~$+~\tilde{h}/\tilde{s}$ theory.  The shaded band 
 corresponds to the variation of $\tilde{m}$ by an order of magnitude 
 above and below $10^{14}~{\rm GeV}$.  The dashed lines show the 
 predictions for $\delta = \pm 0.1$, with $\tilde{m}$ fixed to 
 $10^{14}~{\rm GeV}$.}
\label{fig:M_H-2}
\end{figure}
The sensitivity of the Higgs mass prediction to the scale $\tilde{m}$ 
is shown by the shaded band in Figure~\ref{fig:M_H-2}, where $\tilde{m}$ 
is varied by an order of magnitude above and below $10^{14}~{\rm GeV}$. 
Including all leading-log threshold corrections from an arbitrary 
superpartner spectrum in our theory, we find that just a few of 
the superpartner masses dominate the matching scale
\begin{equation}
  \tilde{m} \simeq 
  \left\{ \begin{array}{ll}
    m_\lambda^{1.6}/m_{\tilde{t}}^{0.6} & \mbox{for small }y
\\ 
    \sqrt{m_s m_H} & \mbox{for large }y
  \end{array} \right.,
\label{eq:mtilde}
\end{equation}
where $m_\lambda$ and $m_{\tilde{t}}$ are the gaugino and top squark 
masses, while $m_s$ and $m_H$ are the masses of the scalar superpartner 
of $\tilde{s}$ and the heavy Higgs boson, respectively.  We ignore 
non-degeneracies amongst the gaugino masses at this high scale.  Although 
$\tilde{m}$ does not coincide with any particular superparticle mass, 
it is in the vicinity of $m_\lambda, m_{\tilde{t}}, m_s$, and $m_H$. 
In the small $y$ region the sensitivity of the Higgs mass to $\tilde{m}$ 
is extremely small, due to a combination of the convergence property 
discussed below and the flatness of the trajectories of $\lambda$ at 
high scales.  This sensitivity to $\tilde{m}$ grows with $y$, as the 
trajectories of $\lambda$ at high scales become less flat, so that 
by $y = 0.8$, varying $\tilde{m}$ by an order of magnitude causes 
a $2~{\rm GeV}$ shift in the Higgs mass prediction.

The sensitivity of the Higgs mass prediction to $\delta$ is also shown 
in Figure~\ref{fig:M_H-2}.  The dashed lines show the predictions for 
$\delta = \pm 0.1$, with $\tilde{m}$ fixed to $10^{14}~{\rm GeV}$.  In 
the small $y$ region there is an important suppression of the correction 
from $\delta$: for $\delta$ of $10\%$ the Higgs mass changes by only about 
$1\%$.  This is due to the convergence property of the RG trajectories 
of $\lambda$, discussed in \cite{Hall:2009nd}.  Remarkably, in the large 
$y$ region, this convergence behavior is much stronger so that, by $y=0.8$, 
$\delta$ becomes irrelevant and one begins to lose sensitivity to the 
value of the supersymmetric boundary condition itself.

\begin{figure}[t]
  \center{\includegraphics[scale=0.97]{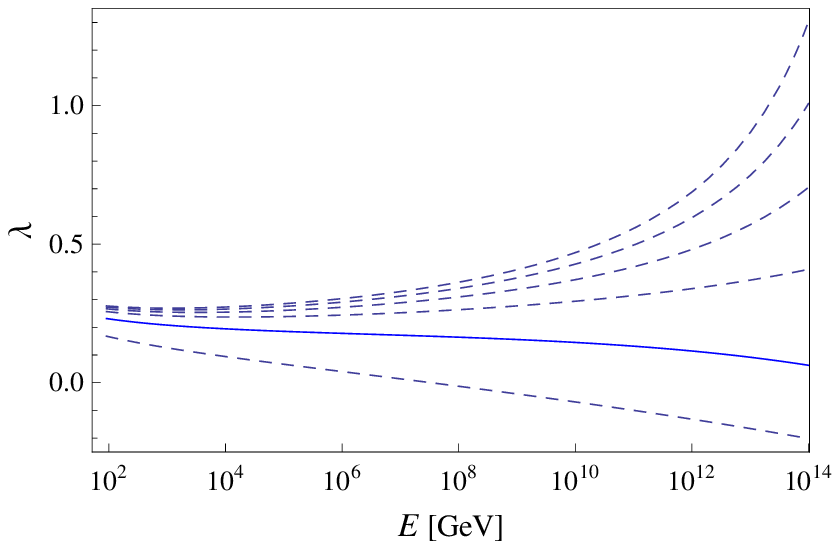}
  \hspace{0.5cm}
  \includegraphics[scale=0.97]{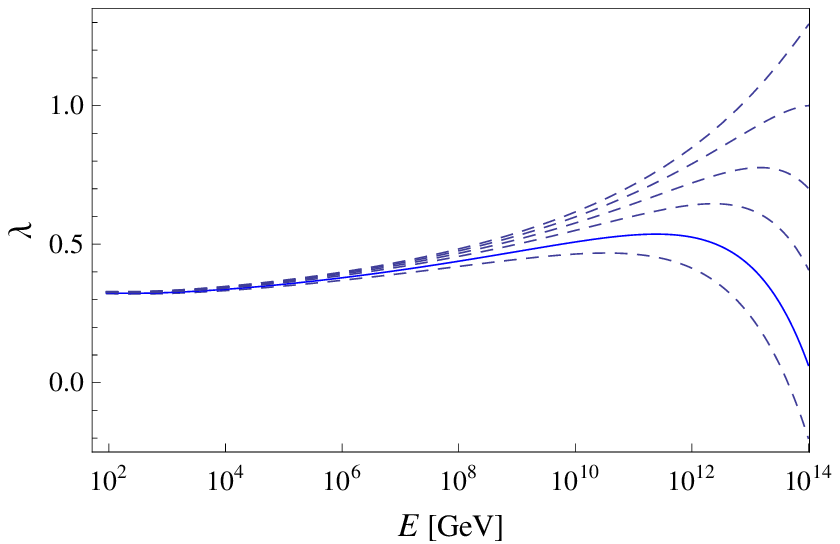}} \\
\caption{Evolution of the Higgs quartic coupling $\lambda$ for $y(M_Z) 
 = 0.7$ (left panel) and $0.9$ (right panel).  The curves represent 
 $\lambda(\tilde{m}) = 0.06$ (the supersymmetric value; solid), $-0.2, 
 0.4, 0.7, 1.0, 1.3$ (dashed), where $\tilde{m} = 10^{14}~{\rm GeV}$. 
 A strong infrared convergence behavior eliminates the sensitivity of 
 the Higgs boson mass to $\lambda(\tilde{m})$ for $y(M_Z) = 0.9$.}
\label{fig:lambda}
\end{figure}
To understand this behavior, in Figure~\ref{fig:lambda} we plot several 
trajectories for $\lambda$, with $y(M_Z) = 0.7$ and $0.9$ for a fairly 
wide range of $\lambda(\tilde{m})$.  Since $y$ grows in the ultraviolet 
it is significantly above unity near $\tilde{m}$, and the $y^4$ term 
in the RG equation for $\lambda$ causes the quartic to rise rapidly on 
scaling below $\tilde{m}$.  Thus $\lambda$ quickly loses its initial 
value and is largely determined by $y$.  Indeed, we see that in much 
of the large $y$ region the Higgs mass is linear in $y$, corresponding 
to $\lambda \propto y^2$ at the weak scale, reflecting the structure 
of the leading terms in the RG equation for $\lambda$.  Therefore, the 
Higgs mass prediction at large $y$ is, in fact, more general than the 
supersymmetric boundary condition.  It arises in any theory where the 
SM is augmented by a weak scale vector-like lepton doublet and Majorana 
singlet, providing the initial value of the quartic coupling is not 
too large and one of the two new gauge invariant Yukawa couplings 
is somewhat small.

In our model, because of the extreme insensitivity to $\lambda(\tilde{m})$ 
discussed above, we need not compute $\delta$ at large $y$.  This is 
fortunate, because in this region there is a large contribution to 
$\delta$ proportional to $y^4$, and $y$ is greater than unity at the 
high scale.  For the small $y$ region this contribution is suppressed, 
and the other contributions to $\delta$ are the same as arise when the 
low energy theory is the SM, and have been computed and discussed in 
detail in~\cite{Hall:2009nd}.  In particular, there is a contribution 
from top squark loops that depends on the dimensionless trilinear 
coupling parameter $A_t/m_{\tilde{t}}$, giving $\delta_{\tilde{t}} 
\sim 0.01~\mbox{--}~0.03$ for $A_t/m_{\tilde{t}} = 1~\mbox{--}~3$. 
In addition, with seesaw neutrino masses there could be a contribution, 
$\delta_\nu$, if the neutrino Yukawa coupling is of order unity.  This 
is significant only if a right-handed neutrino mass, $M_R$, satisfies 
$M_R \simgt 10^{14}~{\rm GeV}$ and $M_R < \tilde{m}$, in which case 
the Higgs mass prediction could be affected at the $1~{\rm GeV}$ level. 
Finally, it may be that threshold corrections from grand unification 
are present, but again we stress that a $10\%$ correction to $\delta$ 
only affects the Higgs mass at the $1\%$ level.

\section{Dark Matter}
\label{sec:DM}

In this section, we study constraints imposed on the parameter space 
by the requirement of dark matter.  We assume the standard picture 
of inflation, followed by a radiation dominated era with high temperature. 
The dark matter particles are produced in the thermal plasma, and their 
abundance is determined by a standard thermal freezeout computation.

The singlino and the neutral Higgsinos mix to form neutralinos, and 
their mass matrix can be read off from the Lagrangian of Eq.~(\ref{eq:L}). 
In the $(\tilde{s}, \tilde{h}_d^0, \tilde{h}_u^0)$ basis, the matrix 
is given by
\begin{equation}
  M_{\chi^0} = \left( \begin{array}{ccc}
    m  & yv   & 0    \\
    yv & 0    & -\mu \\
    0  & -\mu & 0 
  \end{array} \right).
\label{eq:M-neutralino}
\end{equation}
The lightest mass eigenstate $\chi^0_1$ is identified with dark matter. 
In addition to the three neutralinos, there is also one chargino state 
from $\tilde{h}_u^+$ and $\tilde{h}_d^-$, with a mass $\mu$.  Once the 
neutralino mass matrix is diagonalized, we can express the Lagrangian 
in the mass basis and compute all the relevant couplings.  In the 
following subsections, we study constraints on the parameters $y$, 
$m$ and $\mu$ from relic density and direct detection of dark matter.

\subsection{Relic abundance}
\label{subsec:relic-abundance}

The thermally-averaged dark matter annihilation cross section times the 
velocity may be expanded as $\langle \sigma v \rangle = a + b v^2$ for 
small velocities.  The first term is the $s$-wave contribution, while 
the second term the $p$-wave contribution.  The channels contributing 
to the $s$ and $p$-wave annihilations in our model are shown in 
Tables~\ref{tab:s-wave} and \ref{tab:p-wave}.
\begin{table}[t]
\begin{center}
\begin{tabular}{|l||c|}
\hline
$s$-wave  &  channel \\
\hline \hline
$\chi_1^0 \chi_1^0 \rightarrow Z Z$ & 
$t$-channel $\chi_i^0$ ($i=1,2,3$) exchange \\
\hline
$\chi_1^0 \chi_1^0 \rightarrow W^+ W^-$ & 
$t$-channel chargino exchange \\
\hline
$\chi_1^0 \chi_1^0 \rightarrow f \bar{f}$ & 
$s$-channel $Z$ exchange and $s$-channel Higgs exchange \\
\hline
$\chi_1^0 \chi_1^0 \rightarrow h Z$ & 
$t$-channel $\chi_i^0$ ($i=1,2,3$) exchange \\
\hline
\end{tabular}
\end{center}
\caption{Annihilation channels for $s$ wave.}
\label{tab:s-wave}
\end{table}
\begin{table}[t]
\begin{center}
\begin{tabular}{|l||c|}
\hline
$p$-wave  &  channel \\
\hline \hline
$\chi_1^0 \chi_1^0 \rightarrow Z Z$ & 
$s$-channel Higgs exchange \\
\hline
$\chi_1^0 \chi_1^0 \rightarrow W^+ W^-$ & 
$s$-channel Higgs exchange and $s$-channel $Z$ exchange \\
\hline
$\chi_1^0 \chi_1^0 \rightarrow h h$ & 
$s$-channel Higgs exchange and $t$-channel $\chi_i^0$ ($i=1,2,3$) exchange \\
\hline
$\chi_1^0 \chi_1^0 \rightarrow h Z$ & 
$s$-channel $Z$ exchange \\
\hline
\end{tabular}
\end{center}
\caption{Annihilation channels for $p$ wave.}
\label{tab:p-wave}
\end{table}

The cross sections for some of these channels depend on the Higgs 
boson mass, which depends on the Yukawa coupling $y$.  To incorporate 
this dependence we fit the curve in Figure~\ref{fig:M_H-1} to get 
a functional form for the $M_H(y)$.  Since the allowed range in $y$ 
is fairly small, a polynomial provides a very good fit.  Co-annihilation 
contributions become significant once the neutralino masses become 
comparable (when $\mu \sim m$), and we perform our calculation using 
micrOMEGAs~\cite{Belanger:2007zz} to incorporate this effect.

\begin{figure}[]
  \center{\includegraphics[scale=0.9]{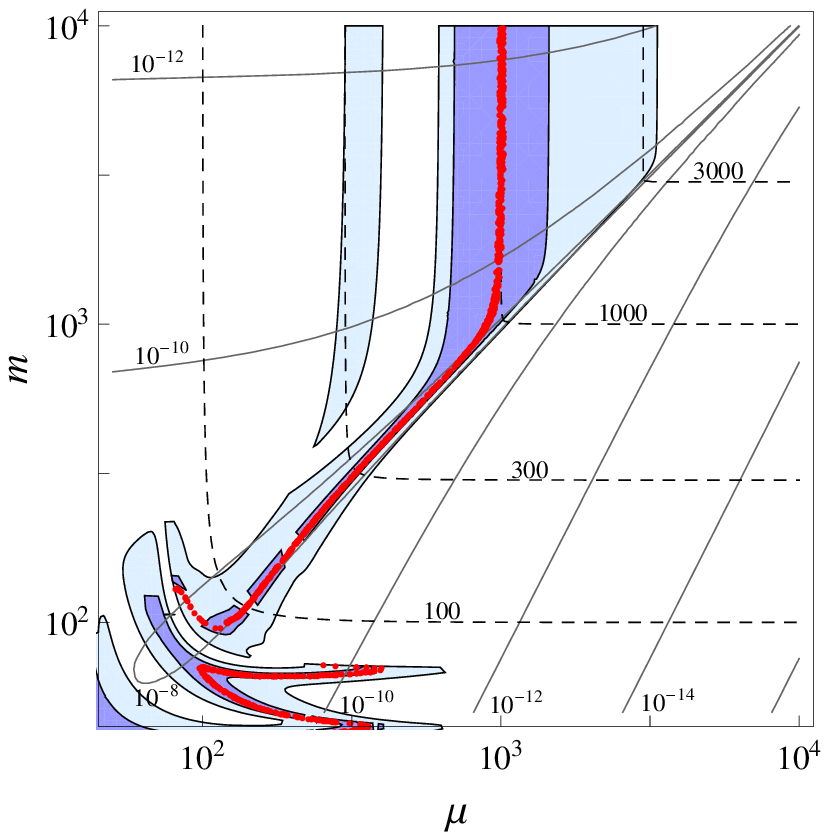}
  \hspace{1.0cm}
  \includegraphics[scale=0.9]{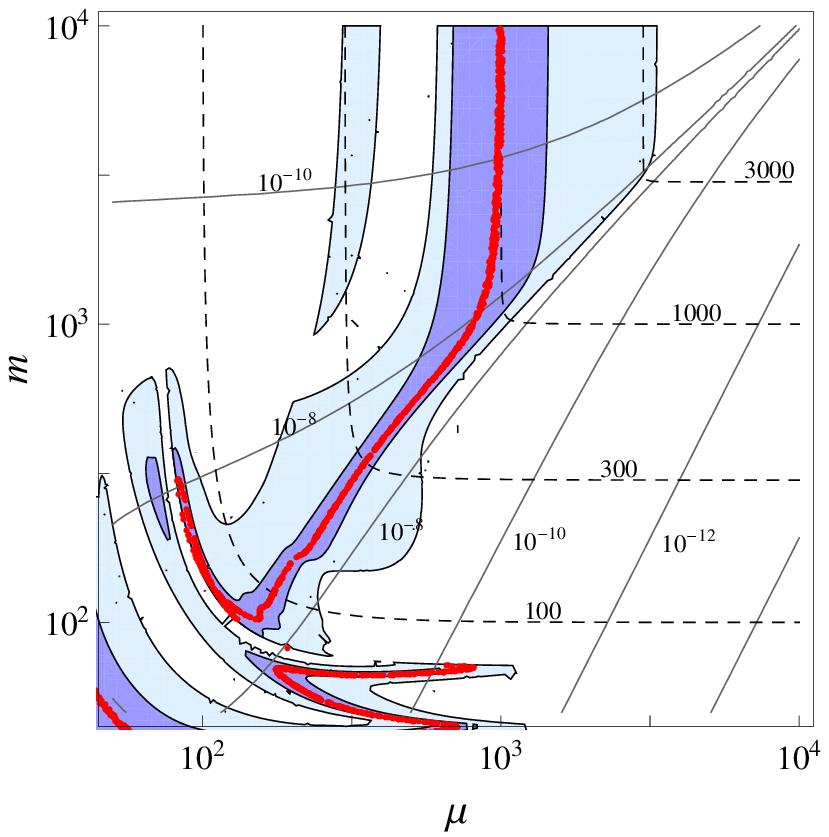}} \\
\vspace{0.5cm}
  \center{\includegraphics[scale=0.9]{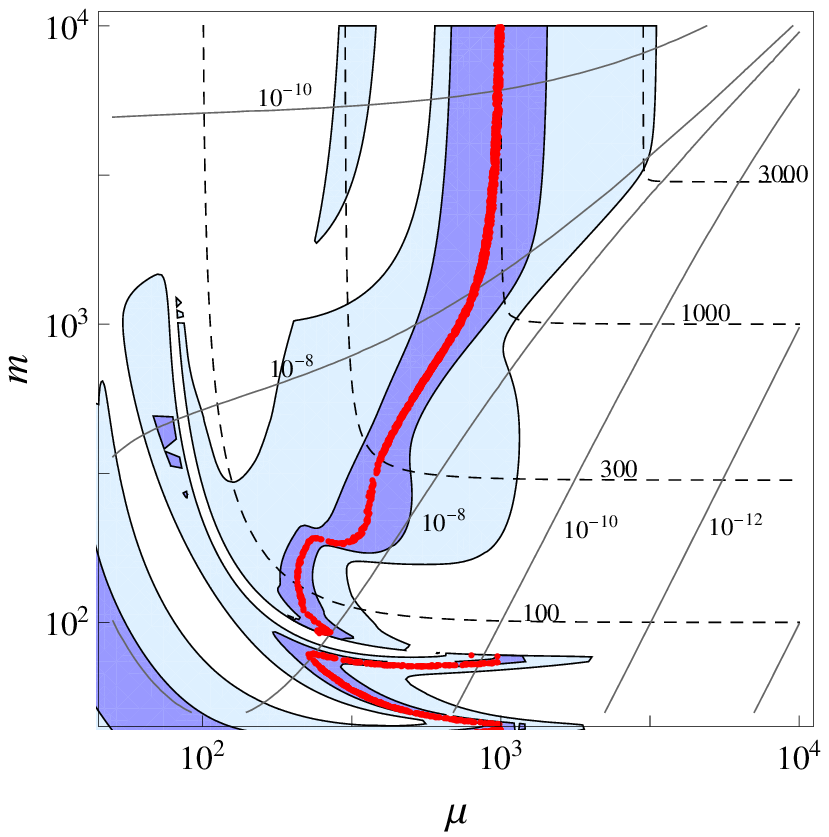}
  \hspace{1.0cm}
  \includegraphics[scale=0.9]{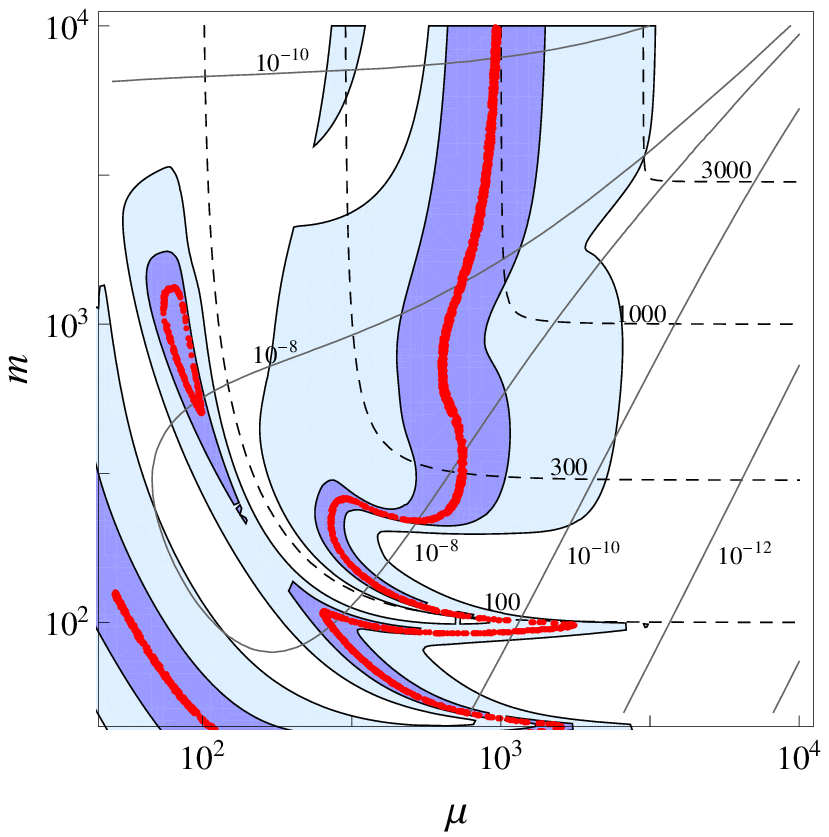}}
\caption{Regions reproducing the observed dark matter abundance 
 on the $\log \mu$-$\log m$ planes (red) for $y=0.2$ (upper left), 
 $0.4$ (upper right), $0.6$ (lower left), and $0.9$ (lower right). 
 The ``anthropic'' windows are also shown in each plot, representing 
 the regions in which the dark matter abundance is within a factor of 
 $2$ (medium blue) and $10$ (light blue) from the central value of 
 observations.  The relic abundance is calculated using micrOMEGAs, 
 which does not include the Sommerfeld effect for annihilation. 
 Inclusion of this effect will shift the values of $\mu$ for the 
 vertical, mostly-Higgsino corridors larger by $\mbox{a few}~\%$. 
 Contours for the dark matter mass (in GeV) are also shown as dashed 
 curves, while those for the spin-independent cross section between 
 dark matter and a nucleon (in pb) are shown as solid curves.}
\label{fig:DM-relic}
\end{figure}
In Figure~\ref{fig:DM-relic}, we show the regions in $\mu$-$m$ planes 
in which the dark matter abundance is consistent with the observed value, 
$\Omega_{\rm DM} h^2 = 0.1131 \pm 0.0034$~\cite{Komatsu:2008hk-2}, for 
$y=0.2$, $0.4$, $0.6$, and $0.9$ (in red).  Contours for the dark matter 
mass are shown as dashed curves.  In the plots, we also show ``anthropic'' 
windows, which correspond to the regions where the dark matter relic 
abundance is within a factor of $2$ (medium blue) and $10$ (light blue) 
from the central value of observations.  The choice we made for the 
widths of the anthropic windows is arbitrary, but it illustrates some 
generic situation in environmental selection for dark matter in our model.

Suppose that the masses of the singlino and Higgsinos, $m$ and $\mu$, 
scan in the multiverse, with the distributions more or less flat in 
logarithms.  In this case the distribution function is approximately 
flat in the $\mu$-$m$ planes of Figure~\ref{fig:DM-relic}, so that 
for a given $y$, the probability of finding $m$ and $\mu$ in a certain 
region is proportional to the area of the anthropic window there. 
We then find from Figure~\ref{fig:DM-relic} that for $y \simlt 0.2$, 
we most likely find $\mu \sim {\rm TeV}$ and $\tilde{m} \simgt {\rm TeV}$, 
as the anthropic window has the largest area there.  This is the region 
in which dark matter is mostly a pure Higgsino, of mass $\sim {\rm TeV}$, 
with a small mixture with the singlino.  (The plots are obtained using 
micrOMEGAs, which does not include the effect of Sommerfeld enhancement 
for annihilation.  Inclusion of this effect will shift the values of 
$\mu$ larger in this region by $\mbox{a few}~\%$~\cite{Hisano:2006nn}.) 
On the other hand, for $y \simgt 0.6$, we find that regions with small 
dark matter masses become larger, so that the probability of finding 
ourselves in one of these regions is significant.  In these regions, 
dark matter is a mixture of the singlino and Higgsinos, and has a mass 
below $\sim 300~{\rm GeV}$.  At what value of $y$ the transition of 
the two regimes occurs?  This depends on the width of the anthropic 
window.  For a relatively small window, having a light dark matter 
regime with a significant probability requires rather large $y$, e.g.\ 
$\simgt 0.6$, as indicated by the medium blue regions in the figure. 
For a large window, however, there is a significant probability of 
obtaining light dark matter already for $y \sim 0.4$, as can be seen 
by looking at the light blue regions in the figure.

In the case that only the symmetry breaking parameter $\epsilon$ scans 
in the multiverse, the ratio between $m$ and $\mu$ is fixed.  In this case, 
scanning occurs along a line with a fixed slope in the $\mu$-$m$ plane. 
Assuming that the distribution of $\epsilon$ is approximately flat in 
logarithm, the situation is almost the same as above for $m/\mu \simgt 1$. 
For $m/\mu \simlt 1$, however, the mostly Higgsino case cannot be realized, 
and light dark matter regions result.

Phenomenology of our model depends crucially on the parameter region we 
are in.  As we saw above, this depends on the assumption on the multiverse 
distribution of parameters, including which parameters do or do not scan, 
as well as the width of the anthropic window for dark matter abundance. 
The relative preference between the light and Higgsino dark matter 
regions is also changed if the distribution of $m$ or $\mu$ deviates 
from logarithms.  We therefore consider both these cases when we analyze 
phenomenology of the model in later sections.

\subsection{Direct detection}
\label{subsec:dir-det}

In addition to the bound from its abundance, dark matter is also 
subject to experimental constraints from direct detection searches. 
In Figure~\ref{fig:DM-relic}, we show contours of the spin-independent 
cross section between dark matter and a nucleon.  We find that the 
current upper bound from existing experiments permits the entire parameter 
space shown.  This can be understood as follows.  Since the dark matter 
particle is in general a mixture of the neutral components of the 
Higgsinos and the singlino, it is a Majorana fermion, so that its 
coupling to the $Z$ boson is suppressed.  The dominant contribution 
to the cross section then comes from Higgs exchange.  In most of 
the parameter region, this contribution is suppressed by the mixing 
between the Higgsino and singlino components, and hence small. 
For $|\mu - m| \sim 100~{\rm GeV}$ the mixing can be large, but the 
cross section is still small enough to evade the current bound.

In the future, experiments such as XENON, LUX and LZ20 are expected to 
lower the upper bound on the cross section and will probe a significant 
portion of the parameter space of the model.  This will be discussed 
in more detail in section~\ref{subsec:DM-direct}.

\section{Experimental Signals}
\label{sec:signals}

In this section, we discuss potential experimental signals which could 
probe the model in the near future.  We mainly consider experiments 
which are either already taking data or will start taking data ``soon.'' 
These include measurements at the Tevatron and the LHC, dark matter 
direct detection experiments like XENON, LUX, LZ20, etc., and indirect 
detection experiments like HESS, VERITAS, MAGIC, FERMI/GLAST, etc. 
We now look at each of these in some detail.

\subsection{Higgs mass measurements at the Tevatron and the LHC}
\label{subsec:higgs-measure}

From Figure~\ref{fig:M_H-1}, we see that the Higgs boson mass ranges 
from $\simeq 141~{\rm GeV}$ to $\approx 210~{\rm GeV}$ for the entire 
perturbatively allowed range of $y$.  For this range of the Higgs 
boson mass, the production cross section at the Tevatron---coming from 
gluon fusion ($gg$), associated production ($Wh,Zh$), and vector boson 
fusion (VBF)---is sizable, ranging from $\sim 830~{\rm fb}$ to $\sim 
195~{\rm fb}$~\cite{CDFNote}.  By combining results from all available 
CDF and D0 channels on SM-like Higgs searches with luminosities 
$2.0~\mbox{--}~4.8{\rm fb}^{-1}$ and $2.1~\mbox{--}~5.4~{\rm fb}^{-1}$, 
respectively, the Tevatron has already excluded an SM-like Higgs 
boson at $95\%~{\rm C.L.}$ in the mass range $163~{\rm GeV} < M_H 
< 166~{\rm GeV}$~\cite{Collaboration:2009je}.  With more data and 
future improvements on the analysis, the Tevatron will be able to 
probe a wider range for the Higgs mass around this region.

A real discovery of the Higgs boson, however, will have to wait the LHC. 
In fact, the LHC can cover the entire Higgs mass range of the model 
relatively easily.  This can be seen, for example, from Figure~13 of 
Ref.~\cite{Abdullin:2005yn} where it is shown that the $h \rightarrow WW$ 
and $h \rightarrow ZZ$ channels can be used to make a $5\sigma$ discovery 
of a SM-like Higgs boson in the above mass range with about $10~{\rm fb}^{-1}$ 
of data.  The best channel for the determination of the Higgs mass in 
this mass range is $h \rightarrow ZZ \rightarrow 4l$ with one of the $Z$s 
being real or virtual depending on the Higgs mass.  It has been claimed 
in~\cite{Ball:2007zza,ATLAS:1999fr} that the statistical precision on 
the Higgs mass at CMS and ATLAS with $30~{\rm fb}^{-1}$ data at the LHC 
is $\sim 0.1\%$.  The systematic uncertainty on the absolute energy scale 
for leptons is expected to be around the same or less; for example, the 
goal of ATLAS is to determine the lepton energy scale to $0.02\%$.  Hence, 
the Higgs mass can be determined very precisely in the mass range relevant 
here.  This is, in fact, crucial in making an accurate determination of $y$ 
from the Higgs boson mass, which can then be tested by other measurements.

\subsection{LHC signals}
\label{subsec:LHC}

We now discuss prospects for probing our model at the LHC.  The possible 
production modes for the new TeV states are $q\bar{q} \rightarrow \chi_1^+ 
\chi_1^-,\, \chi_i^0 \chi_j^0$; $q\bar{q}' \rightarrow \chi_1^+ \chi_i^0$; 
and $gg \rightarrow \chi_i^0\chi_j^0$ through an intermediate on-shell 
Higgs boson produced via top-loop.  The possible decay modes are $\chi_3^0 
\rightarrow \chi_{1,2}^0 Z,\, \chi_1^{\pm}\,W^{\mp}$; $\chi_1^{\pm} 
\rightarrow \chi_1^0 W^{\pm}$; and $\chi_2^0 \rightarrow \chi_1^0 Z$. 
The signatures are, therefore, based on ``short cascades'' in contrast 
to the ``long cascades'' in traditional supersymmetry.  The best signals 
arise when the $W$ and $Z$ bosons decay leptonically giving rise to 
various numbers of leptons and missing energy.  The hadronic channels 
suffer from larger backgrounds, but a larger branching ratio of the weak 
bosons to hadrons suggests that a detailed investigation may be necessary. 
Here we focus on the leptonic channels.  The signatures are similar to 
those studied in~\cite{Enberg:2007rp}, except that the Higgs mass was 
quite heavy in the model considered there.

From the analysis in Ref.~\cite{Enberg:2007rp}, one finds that even if 
one restricts to leptonic channels, those with a) 1~lepton + $\notEt$ 
and b) 2~leptons + $\notEt$\, do not provide good prospects.  This 
is because of the huge SM backgrounds.  For channel a), the dominant 
backgrounds are Drell-Yan $W^{(*)}$ production, $WZ$ production and 
$t\bar{t}$ production, and they swamp the signal even for parameter 
regions with large cross sections.  For channel b), which arises from the 
production of $\chi_i^0 \chi_1^0$ and $\chi_1^+ \chi_1^-$, the dominant 
backgrounds are from $WW$, $ZZ$, $WZ$ and $t\bar{t}$ production.  By 
looking at the $H_T$ ($\equiv \sum_{\rm visible}|p_T|$) distribution 
after applying appropriate cuts, one finds that the background and 
signal are very similar in shape.  For large cross sections of $\sim 
200~{\rm fb}$, it may be possible to get a good signal significance 
after optimizing the cuts for about $100~{\rm fb}^{-1}$ of data. 
In Ref.~\cite{Enberg:2007rp}, however, the large enhancement in the 
production cross section for $\chi_i^0 \chi_1^0$ arose from the Higgs 
production mode, with $h \rightarrow \chi_i^0 \chi_1^0$ for a heavy 
enough Higgs boson.  This mode is not available in our model, so 
channel b) does not provide good prospects either.

Channel c) with 3~leptons + $\notEt$\, provides better prospects. 
Search strategy in this channel is similar to that for direct 
chargino-neutralino production in standard supersymmetry, with some 
differences.  Analyses for the mSUGRA model for the trilepton channel 
have been performed in detail in~\cite{Aad:2009wy} for ATLAS and 
in~\cite{Ball:2007zza} for CMS.  A more general MSSM model has been 
studied in~\cite{Vandelli:2007zz} which is better suited for comparing 
to our model.  In the limit where the squarks and sleptons are heavy, 
$\chi_1^+ \chi_{2,3}^0$ production proceeds through the same diagrams 
both for the MSSM and our model, with the dominant one being Drell-Yan 
production via $s$-channel $W$ exchange.  The dominant SM backgrounds 
are from $t\bar{t}$ production with both tops decaying leptonically and 
the third lepton coming from a semileptonic decay of the $b$ or a jet 
fake, and $WZ$ production with both $W$ and $Z$ decaying leptonically. 
In the MSSM, there is also a large supersymmetry background from gluino 
pair production and gluino squark production.  If gluino and squarks 
are much heavier than the charginos and neutralinos, these supersymmetry 
backgrounds can be minimized.  Also, in these analyses the parameter 
region $m_{\chi_2^0} - m_{\chi_1^0} < M_Z$, in which an opposite sign, 
same flavor (OSSF) pair of leptons arises from an off-shell $Z$, is mostly 
studied as this is quite helpful in reducing the $WZ$ background in the 
dilepton mass distribution where the $Z$ is on-shell and decays to two 
OSSF leptons.  Apart from the $Z$-mass cut, the other important cuts 
applied in~\cite{Aad:2009wy,Ball:2007zza,Vandelli:2007zz} are: 1) At least 
three isolated leptons, 2) One pair of OSSF leptons, and 3) No jets with 
$p_T > 30~{\rm GeV}$ (optional).  Figure~4.17 in~\cite{Vandelli:2007zz} 
shows the $5\sigma$ sensitivities with $30~{\rm fb}^{-1}$ and 
$100~{\rm fb}^{-1}$ of data at the LHC in the $\mu$-$M_2$ plane 
(with $M_2 = 2M_1$), where $\mu$ is the Higgsino mass and $M_2$, 
$M_1$ are the wino and bino mass parameters in the MSSM.

\begin{figure}[]
  \center{\includegraphics[scale=0.9]{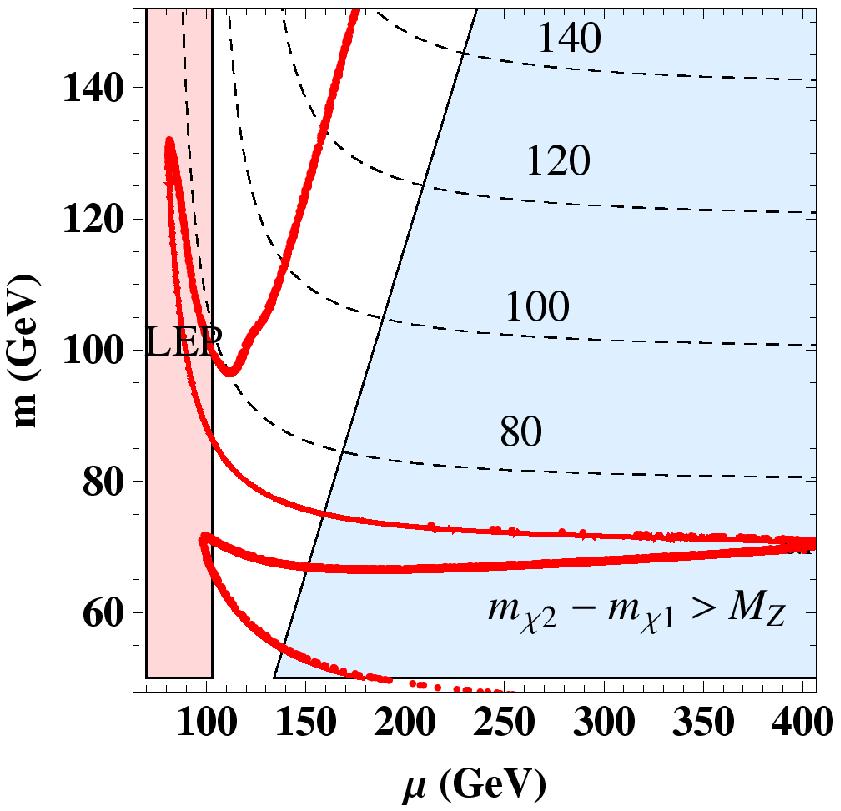}
  \hspace{1.0cm}
  \includegraphics[scale=0.9]{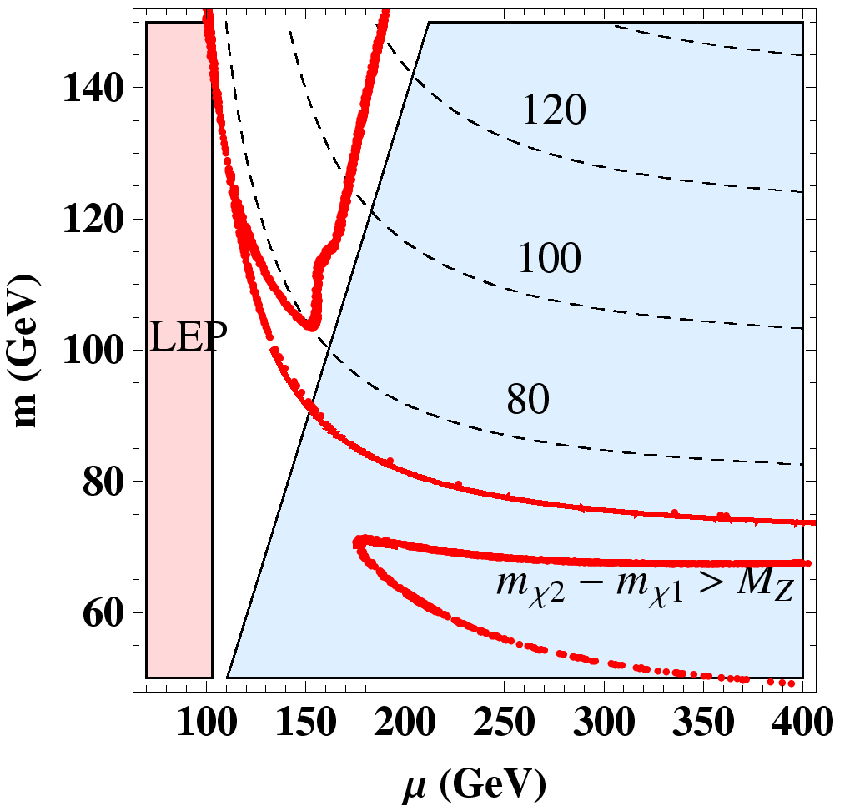}} \\
\vspace{0.5cm}
  \center{\includegraphics[scale=0.9]{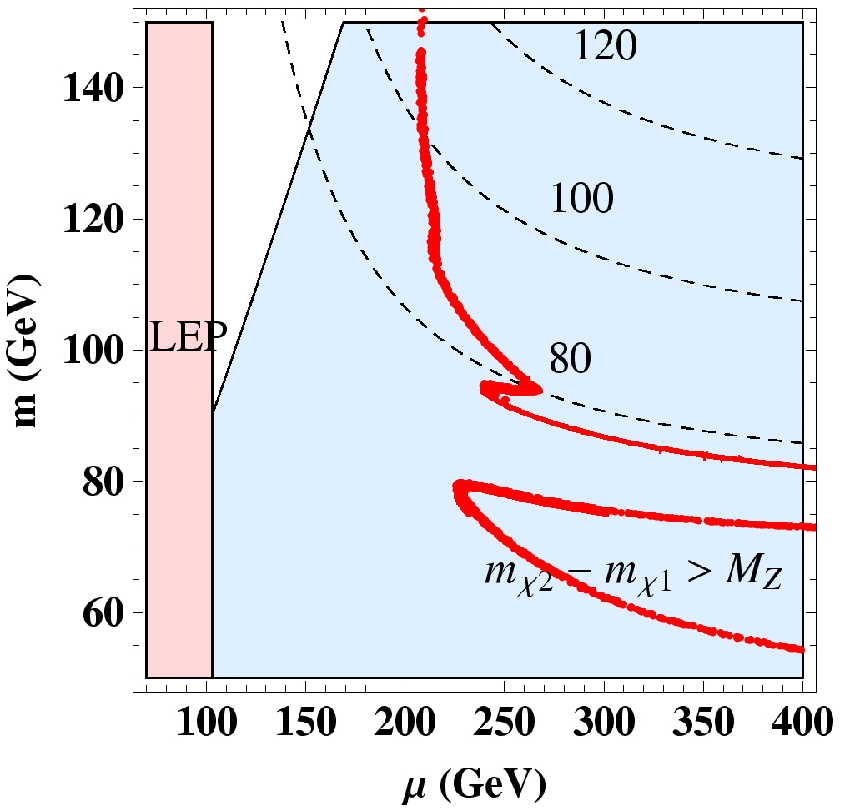}
  \hspace{1.0cm}
  \includegraphics[scale=0.9]{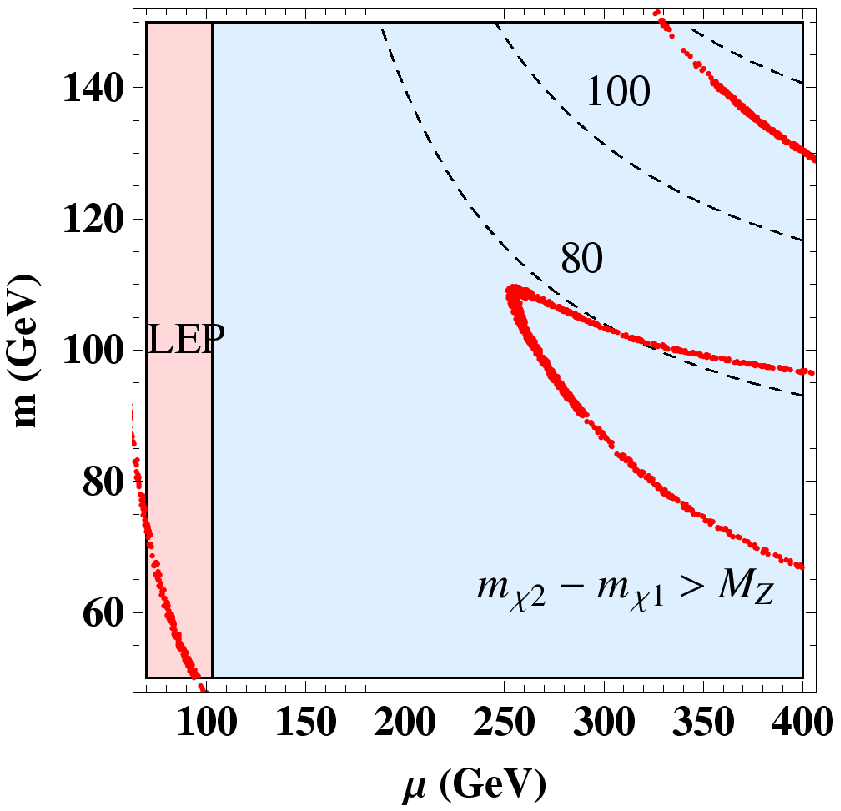}}
\caption{Regions consistent with the LEP constraint on the 
 chargino mass, $m_{\chi_1^+} > 103~{\rm GeV}$, and the $Z$-mass cut, 
 $m_{\chi_2^0} - m_{\chi_1^0} < M_Z$, on the $\mu$-$m$ planes (unshaded 
 regions) for $y=0.2$ (upper left), $0.4$ (upper right), $0.6$ (lower 
 left), and $0.9$ (lower right).  The red bands correspond to the 
 parameter regions consistent with the WMAP bound on the dark matter 
 relic abundance.  Contours for the dark matter mass (in GeV) are 
 also drawn as dashed curves.}
\label{fig:5-sigma}
\end{figure}
What does this imply for our model?  Since the production and decay 
modes for the channels considered here are identical between our 
model and the MSSM, one expects that the production cross section 
times branching ratio ($\sigma \times {\rm Br}$) is also roughly 
similar.  This implies that for channel c), the regions with cross 
sections $\sim 10~\mbox{--}~50~{\rm fb}$ could be probed in our 
model.  In Figure~\ref{fig:5-sigma}, the regions in the $\mu$-$m$ plane 
which are consistent with i) the LEP constraint on the chargino mass 
($m_{\chi_1^+} > 103~{\rm GeV}$~\cite{Amsler:2008zzb}) and ii) the 
$Z$-mass cut ($m_{\chi_2^0} - m_{\chi_1^0} < M_Z$) are shown in 
white (unshaded), for four different values of $y$.  The regions iii) 
leading to the WMAP allowed relic abundance~\cite{Komatsu:2008hk-2} 
are also depicted with red bands.  The unshaded regions within the 
range of the plots lead to production cross sections of roughly 
$O(10~\mbox{--}~50~{\rm fb})$, and hence should be probed with around 
$100~{\rm fb}^{-1}$ of data, using the method similar to that described 
above.  We then find that for small $y \simlt 0.4$, the region 
$103~{\rm GeV} < \mu \simlt (130~\mbox{--}~200)~{\rm GeV}$ and 
$50~{\rm GeV} \simlt m \simlt (140~\mbox{--}~150)~{\rm GeV}$, which 
corresponds to $m_{\chi_1^0} \simlt (120~\mbox{--}~140)~{\rm GeV}$, 
can be probed at the LHC with $100~{\rm fb}^{-1}$ of data.  For larger 
values of $y$, however, the region consistent with i) and ii) shrinks 
and the WMAP constraint iii) is not satisfied.  Thus, it is very 
hard to probe these regions at the LHC with analysis similar 
to~\cite{Aad:2009wy,Ball:2007zza,Vandelli:2007zz}, even with 
large amount of data.

It is also important to know if precise measurements of the model 
parameters can be made for the parameter regions that can be probed 
at the LHC.  An observation of the dilepton invariant mass edge from 
OSSF leptons arising from an off-shell $Z$ boson can, in principle, 
determine the mass difference $m_{\chi_{2,3}^0} - m_{\chi_1^0}$. 
Since the systematic uncertainty on the lepton energy scale is quite 
small ($\simlt 0.1\%$), the uncertainty is dominated by statistics. 
It turns out that with a cross section of $10~\mbox{--}~50~{\rm fb}$ 
the number of signal events after imposing cuts with $100~{\rm fb}^{-1}$ 
of data is only $O(100)$, and is about $2.5~\mbox{--}~3$ times smaller 
than the SM background~\cite{Vandelli:2007zz}.  Thus, the statistical 
precision of the measurement is limited by the bin size, which has to 
be large to contain enough events such that they are sufficiently above 
the fluctuation of the background.  A rough estimate from the analysis 
in~\cite{Vandelli:2007zz} gives the precision to be around $10\%$. 
Of course, the precision can be improved with more data, but given 
that the statistical precision with $100~{\rm fb}^{-1}$ of data is 
about $10\%$, the precision can only be increased by a factor of few. 
A more detailed analysis is needed to obtain a better estimate of 
the precision, however.

\subsection{Direct detection of dark matter}
\label{subsec:DM-direct}

Since the reach of the LHC is only limited to small dark matter masses 
($m_{\chi_1^0} \simlt 140~{\rm GeV}$) as seen in the previous subsection, 
it is important to look at prospects for probing the model with direct dark 
matter detection experiments.  As we saw in section~\ref{subsec:dir-det}, 
the entire parameter space of our model is allowed with existing bounds. 
This is because the $Z$-exchange contribution is highly suppressed, 
while the Higgs exchange contribution is still small.  The one-loop 
contribution to the cross section generated by gauge interactions, which 
could be important for the mostly Higgsino-like case~\cite{Hisano:2004pv}, 
turns out to be insignificant for model parameters satisfying the WMAP 
relic abundance constraint.

\begin{figure}[t]
  \center{\includegraphics[scale=0.9]{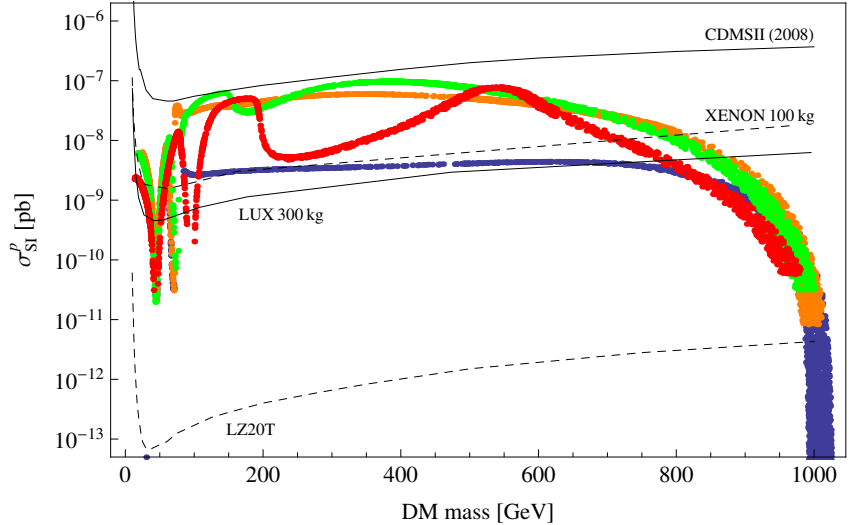}}
\caption{Predictions for the spin-independent cross section of dark 
 matter on a proton, obtained by scanning $\mu$ and $m$ for four different 
 values of $y$: $y=0.1$ (blue), $0.4$ (orange), $0.6$ (green) and $0.9$ 
 (red).  The densities of the dots are the results of the scan, and do 
 not have particular significance.  Also shown are existing bounds from 
 CDMSII and projected $90\%$~C.L. upper limits from XENON~100kg, LUX~300kg, 
 and LZ20T; these curves are obtained using Ref.~\cite{DM-tool}.}
\label{fig:dd-scan}
\end{figure}
In Figure~\ref{fig:dd-scan}, we scan over values of $\mu$ and $m$ allowed 
by the WMAP result on the relic abundance for four different values of 
$y$, and plot the corresponding spin-independent cross section on a proton 
$\sigma_{\rm SI}^p$, calculated using micrOMEGAs.  The scan is performed 
over the region where $m < 10~{\rm TeV}$, motivated by our theoretical 
construction in section~\ref{sec:model}.  It is quite promising that 
the \emph{entire} region of this parameter space will be probed by future 
generations of direct detection experiments.  In particular, dark matter 
masses in the range $100~{\rm GeV} \simlt m_{\chi_1^0} \simlt 700~{\rm GeV}$ 
can be probed relatively soon for the entire range of $y$, by XENON 
and LUX.  To probe heavier dark matter, more advanced detectors are 
required.  For example, to probe the largest dark matter masses around 
a TeV, the highly advanced LZ20T detector, with a sensitivity about 
three orders of magnitude better than that of LUX~300kg, will be needed.

In addition to the detection of dark matter, it is important to address 
how precisely the mass and cross section can be determined from these 
experiments.  This has been studied in a number of papers, the most 
relevant ones being~\cite{Shan:2009ym,Drees:2008bv,Green:2008rd}.  Here 
we follow the results of~\cite{Shan:2009ym}.  It was shown there that 
with data from only one experiment, the dark matter mass combined 
with its spin-independent cross section can be determined by a maximum 
likelihood analysis, comparing a theoretically predicted spectrum to 
measured recoil energies.  For small dark matter masses ($m_{\chi_1^0} 
< m_{\rm nucleus}$), the shape of the recoil spectrum is sensitive 
to the mass, enabling us to determine the mass (as well as the 
cross section) with modest accuracy.  For example, for a dark 
matter mass of $50~{\rm GeV}$ and spin-independent cross section 
of $10^{-7}~{\rm pb}$, the statistical uncertainty for the mass turns 
out to be around $10\%$ with a reasonably large exposure of $3 \times 
10^5~{\rm kg}\cdot{\rm days}$.  The statistical uncertainty worsens to 
about $20\%$ if the mass is raised to $100~{\rm GeV}$ with the same cross 
section and exposure.  For dark matter masses above $100~{\rm GeV}$, 
the uncertainty increases drastically because the recoil spectrum 
is insensitive to the mass.  In addition, with data from only one 
experiment, the mass determination described above depends on the 
assumptions about the velocity profile and also the local dark matter 
density.  For example, for a dark matter mass of $100~{\rm GeV}$, 
a $10\%$ error in the velocity distribution will cause a $20\%$ 
systematic error in the mass determination, with a $\sim 10\%$ 
error in the spin-independent cross section.

With data from two or more experiments, however, it is possible to 
obtain the dark matter mass without any assumption about the velocity 
profiles or the local density.  This is because the normalized 
(one-dimensional) velocity distribution $f_1(v)$ can be solved 
directly from the expression for the direct detection rate, and 
is independent of the spin-independent cross section and the local 
halo density~\cite{Drees:2008bv}.  However, it does depend on the 
masses of the nucleus and the dark matter.  Thus, by requiring that 
the values of a given moment of the distribution $f_1(v)$ agree for 
two experiments with nuclei $X$ and $Y$, it is possible to measure 
the dark matter mass $m_{\chi_1^0}$ in a model-independent way. 
For $m_{\chi_1^0} \simlt 100~{\rm GeV}$, it is possible to measure 
the dark matter mass to $\simlt 20\%$ with $O(500)$ events (before cuts) 
for two target nuclei ${}^{28}\!\,{\rm Si}$ and ${}^{76}\!\,{\rm Ge}$. 
For masses above $100~{\rm GeV}$, however, the algorithmic procedure 
for determining the masses does not provide reliable estimates. 
Finally, by making an assumption about the local dark matter density, 
it is also possible to measure the spin-independent cross section 
with a statistical uncertainty of $\sim 15\%$ with $O(50)$ events 
for $m_{\chi_1^0} \simlt 100~{\rm GeV}$, without any assumption about 
the dark matter mass or its velocity profile.  Again, the uncertainty 
in the measurement of the cross section increases drastically for 
$m_{\chi_1^0} \simgt 100~{\rm GeV}$~\cite{Shan:2009ym}.

\subsection{Indirect detection of dark matter}
\label{subsec:DM-indirect}

In addition to direct detection, dark matter can also be detected 
through products of its annihilation in the galactic halo, which 
gives rise to many kinds of cosmic rays such as photons, neutrinos, 
electrons, positrons, protons and antiprotons.  Since the annihilation 
cross section of dark matter is determined by the measured relic 
abundance, and since there is no significant enhancement of the 
cross section from the epoch of freezeout to the present, the indirect 
detection signals from electrons and positrons, protons and antiprotons, 
and diffuse photons and neutrinos are not very promising as they 
are typically overwhelmed by the background.

However, the processes $\chi_1^0 \chi_1^0 \rightarrow \gamma \gamma$ 
and $\chi_1^0 \chi_1^0 \rightarrow \gamma Z$ are induced at loop level, 
leading to monochromatic $\gamma$-ray with energies $E_{\gamma} = 
m_{\chi_1^0}$ and $E_{\gamma} = m_{\chi_1^0} - M_Z^2/4m_{\chi_1^0}$, 
respectively.  This gives rise to sharp lines in the photon spectrum 
which have much better prospects for detection.  Adapting the detailed 
computation of $\sigma v$ given in~\cite{Bergstrom:1997fj} to our model, 
we find that $\sigma v (\chi_1^0\chi_1^0 \rightarrow \gamma\gamma)$ 
reaches a relatively large constant value ($\approx 10^{-28}~{\rm cm^3/s}$) 
for Higgsino-like dark matter with mass $\simgt~{\rm TeV}$.  A more 
precise treatment of this case should include non-perturbative effects 
involving both dark matter and its $SU(2)_L$ partner(s) (the Sommerfeld 
enhancement)~\cite{Hisano:2003ec}.  However, since including these 
effects changes $\sigma v (\chi_1^0\chi_1^0 \rightarrow \gamma\gamma)$ 
by at most $O(30\%)$ for dark matter masses satisfying the 
relic abundance constraint, we use the perturbative computation 
of~\cite{Bergstrom:1997fj} in our analysis, for simplicity.

The detectors best suited for detecting $\gamma$ lines are atmospheric 
Cherenkov telescope (ACT) detectors like VERITAS, HESS, MAGIC, etc.\ 
and satellite-borne $\gamma$-ray detectors like FERMI/GLAST.  Although 
the ACT detectors have a small angular acceptance and a higher energy 
threshold, they have a much larger area than satellite-based telescopes 
like FERMI/GLAST.  So, they are ideal when a large monochromatic photon 
flux is emitted from a small region of the sky such as from the Galactic 
Center, whose coordinates are known with sufficient accuracy.  This 
also means that they are strongly sensitive to the dark matter profile 
and provide better prospects for steeper profiles like NFW or Moore. 
The sensitivity of ACT detectors is determined by a relatively large 
background of misidentified gamma-like hadronic showers and cosmic-ray 
electrons, with the contribution from the diffuse $\gamma$-ray background 
being mostly subdominant.  MAGIC currently provides the best sensitivity 
for $\gamma$-rays with energies larger than $300~{\rm GeV}$; the 
sensitivities of HESS and VERITAS are very similar.  Advanced 
detectors like AGIS and CTA, which will come online in future, have 
the potential to reach much better sensitivities than the detectors 
above~\cite{Buckley:2008ud}, although their physical location will 
determine their ability to observe the Galactic Center.  The $5\sigma$ 
point-source sensitivities of MAGIC, VERITAS and HESS for continuous 
$\gamma$-rays are given in~\cite{Bergstrom:1997fj,Arvanitaki:2004df}, 
which we adopt in our analysis.  (The sensitivities for the $\gamma$-lines 
are in fact expected to be somewhat better.)  There are, however, two 
important sources of uncertainties in these detectors.  First, these 
detectors have a $10~\mbox{--}~20\%$ energy resolution which smears 
the line shape and decreases the sensitivity.  Second, there is an 
overall systematic uncertainty in the energy scale, $\sim 30\%$ for 
MAGIC and $\sim 15\%$ for HESS, which provides the dominant uncertainty 
in the determination of the dark matter mass.

Unlike ACT detectors, satellite detectors have a much larger angular 
acceptance and a lower energy threshold.  These detectors look at 
a large part of the galaxy; hence they are much less sensitive to 
the dark matter profile.  Therefore, detectors like FERMI/GLAST provide 
the best sensitivity for less steep profiles.  Since the background 
rejection from hadronic showers and cosmic electrons is much better 
for these detectors, their sensitivity is only limited by counting 
statistics and the diffuse $\gamma$-ray background.  The $5\sigma$ 
sensitivity for a five-year operation of FERMI/GLAST is given 
in~\cite{Baltz:2008wd}.

The Galactic Center is ideal for the detection of the gamma-ray monochromatic 
signal because of the extremely large dark matter density there.  However, 
the presence of astrophysical point sources in the Galactic Center could 
present a serious background to the signal.  In fact, an intense point-like 
gamma-ray source (J1745-290) has been observed by HESS~\cite{Aharonian:2006wh}. 
One possible origin of this source is believed to be the stochastic 
acceleration of electrons interacting with the turbulent magnetic field 
of Sgr~A*, the supermassive black hole at the center of our galaxy. 
In the following analysis, for simplicity we compute sensitivities for 
$\sigma v (\chi_1^0\chi_1^0 \rightarrow \gamma\gamma)$ (for different 
profiles) assuming that this important background can be subtracted 
efficiently.  A better (but more complicated) strategy may be to 
consider a larger angular region with a size depending on the dark 
matter profile as well as the morphology of background and signal 
emissions, and then subtract all detected astrophysical sources within 
this region \cite{Serpico:2008ga}.

\begin{figure}[t]
  \center{\includegraphics[scale=0.75]{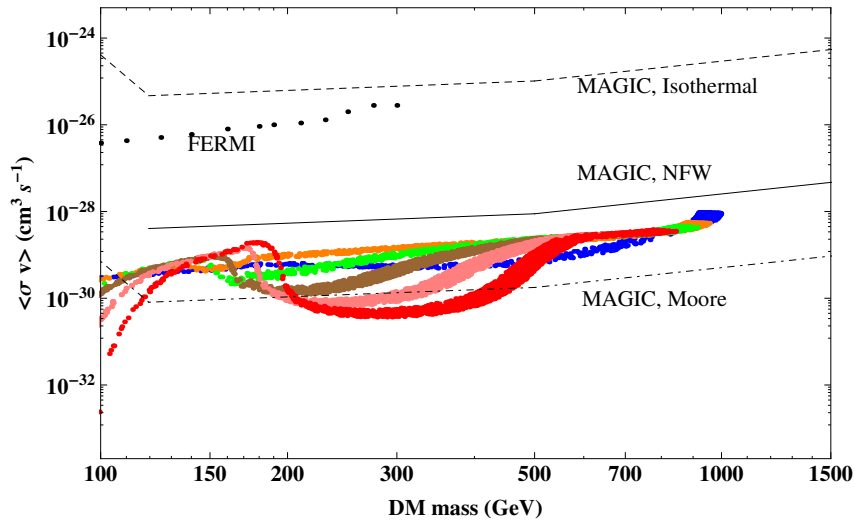}}
\caption{Predictions for $\sigma v (\chi_1^0\chi_1^0 \rightarrow 
 \gamma\gamma)$ by scanning $\mu$ and $m$ for six different values of 
 $y$; the blue, orange, green, pink, brown, and red dots correspond to 
 $y=0.1, 0.4, 0.6, 0.7, 0.8$, and $0.9$, respectively.  Also shown are 
 sensitivities of the MAGIC ACT detector (with one-year exposure) arising 
 from the flux of a $\Delta\Omega = 10^{-5}~{\rm sr}$ cone encompassing 
 the Galactic Center, for three different dark matter profiles: Moore, 
 NFW and Cored Isothermal.  The black dots represent the sensitivity of 
 the FERMI/GLAST experiment, with five-year exposure to the flux of an 
 annulus centered around the Galactic Center, but excluding the region 
 within $15^\circ$ of the Galactic plane.}
\label{fig:photon-scan}
\end{figure}
In Figure~\ref{fig:photon-scan}, we show the results for $\sigma v 
(\chi_1^0\chi_1^0 \rightarrow \gamma\gamma)$ by performing a scan over 
$\mu$ and $m$ (allowed by the WMAP result for the relic abundance) 
for six different values of $y$, similar to what was done for direct 
detection.  The sensitivities of ACT detector---MAGIC with one-year 
exposure to the Galactic Center point source---for three qualitatively 
different dark matter profiles, as well as the sensitivity of FERMI/GLAST 
with five-year exposure to an annulus around the Galactic Center, are 
also shown.  It can be seen that existing ACT detectors like MAGIC, 
HESS or VERITAS provide better detection prospects than FERMI/GLAST for 
reasonably steep profiles.  Still, they can only probe the model for 
very steep halo profiles like the Moore profile.  Regarding these profiles, 
it is interesting to note that the presence of baryons is \emph{not} 
taken into account in the various simulations leading to the above 
profiles.  In principle, this could significantly affect the picture 
near the galactic center because of the interactions of the supermassive 
black hole with the surrounding dark matter distribution.  Although 
a full understanding of these effects is unavailable at present, in the 
``adiabatic compression'' scenario which takes into account some of these 
interaction effects~\cite{Bertone:2005hw}, the dark matter density near 
the galactic center is significantly increased relative to the NFW profile 
leading to a signal which is about three orders of magnitude larger 
than that for the NFW profile~\cite{Bertone:2009cb}!  In this case, 
the entire parameter space of the model can be easily probed with 
the above ACT detectors.

The sensitivities quoted above may be improved if the energy resolution 
is improved or if the background from misidentified photons is further 
suppressed.  Also the $2\gamma$ and $\gamma Z$ lines cannot be resolved 
for dark matter masses $\simgt 500~{\rm GeV}$, in which case adding 
the $\gamma Z$ contribution will increase the effective cross section 
by a little more than a factor of two, making the detection prospects 
better.  Also, as mentioned earlier, future detectors like AGIS and CTA 
are expected to have much better sensitivities and can probe a much larger 
parameter space of the model even for widely accepted profiles such as NFW. 
Moreover, with a better understanding of the systematics, the dark matter 
mass can be measured to better than $\sim 15\%$.

\section{SM~{\boldmath $+$}~Wino as an Effective Theory below the Unified Scale}
\label{sec:wino}

Environmental selection of dark matter may prefer a light wino, so that 
the effective theory below $\tilde{m}$ is SM~$+~\tilde{w}$.  In this 
case the theory at $\tilde{m}$ need not have a singlet---the MSSM 
states are enough to lead to consistent phenomenology at low energies. 
For $\tilde{m}$ not far from the unified scale, $\sim 10^{14}~{\rm GeV}$, 
this setup leads to the following prediction for the Higgs boson 
mass~\cite{Hall:2009nd}.  The Higgs mass $M_H$ is in the range 
$127~{\rm GeV} \simlt M_H \simlt 142~{\rm GeV}$, where we have included 
two-loop RG effects from the wino.  In particular, for large $\tan\beta$
\begin{equation}
  M_H \simeq 141.5~{\rm GeV} 
    - 0.4~{\rm GeV} 
      \left( \frac{10}{\tan\beta} \right)^2 
    + 1.8~{\rm GeV} 
      \left( \frac{m_t - 173.1~{\rm GeV}}{1.3~{\rm GeV}} \right) 
    - 1.0~{\rm GeV} 
      \left( \frac{\alpha_s(M_Z)-0.1176}{0.002} \right).
\label{eq:MH-wino}
\end{equation}
This prediction is subject to errors coming, e.g., from (a few orders of 
magnitude) uncertainty for the value of $\tilde{m}$, supersymmetric threshold 
corrections at $\tilde{m}$, and higher order corrections.  These errors, 
however, give only a few hundred MeV corrections to $M_H$~\cite{Hall:2009nd}, 
so that the prediction is very precise.

The wino $\tilde{w} = (\chi^\pm, \chi^0)$ is an $SU(2)_L$-triplet Weyl 
fermion, and has couplings to the standard model particles only via 
$SU(2)_L$ gauge bosons.  The couplings are completely determined by gauge 
symmetry and therefore the only parameter in the model is the wino mass 
$M_2$.  The charged and the neutral winos are degenerate at tree level. 
After electroweak symmetry breaking, radiative corrections generate a mass 
splitting~\cite{Pierce:1996zz}
\begin{equation}
  \delta m \equiv m_{\chi^\pm} - m_{\chi^0} 
  = \frac{\alpha M_W}{2(1+c_w)} 
    \left( 1 - \frac{3}{8c_w} \frac{M_W^2}{M_2^2} \right),
\label{eq:wino-split}
\end{equation}
which is roughly $165~{\rm MeV}$ for $M_2 \gg M_W$.

Since the only parameter in the model is the wino mass $M_2$, assuming 
dark matter is thermal, $M_2$ can be determined by the observed dark matter 
relic density.  Taking the range of the relic density consistent with 
the observation at $2\sigma$, i.e. $0.106 < \Omega_{\rm DM} h^2 < 0.120$, 
the wino mass is found, using micrOMEGAs, to be $2.24~{\rm TeV} < M_2 
< 2.38~{\rm TeV}$.  As $M_2$ is much larger than the weak boson mass 
$M_W$ and the charged-neutral winos are nearly degenerate, however, 
the non-perturbative Sommerfeld effect can affect the relic density by 
as much as $30\%$.  The effect can be estimated using the numerical result 
given in~\cite{Hisano:2006nn}, where wino-like dark matter was studied. 
The wino mass consistent with the relic density turns out to be
\begin{equation}
  2.7~{\rm TeV} \simlt M_2 \simlt 3.0~{\rm TeV}.
\label{eq:wino-mass}
\end{equation}
This narrow window for $M_2$, as well as the Higgs mass prediction of 
Eq.~(\ref{eq:MH-wino}), are the main predictions of this theory.  To test 
the theory experimentally, therefore, it is important to measure $M_2$ 
with certain precision.

With a large mass of Eq.~(\ref{eq:wino-mass}), discovery of wino dark 
matter at the LHC is not possible.  Dark matter direct detection is also 
hard as the only coupling of dark matter is to the chargino and $W$ gauge 
boson.  This coupling only allows either tree-level inelastic scattering 
or one-loop suppressed elastic scattering with nuclei.  Due to the 
$O(100)~{\rm MeV}$ mass splitting, observation through inelastic scattering 
is not possible.  The dark matter-nucleon elastic scattering cross section 
at one loop is given in Ref.~\cite{Essig:2007az} as about $2 \times 
10^{-9}~{\rm pb}$.  It is beyond the reach of XENON~100kg or LUX~300kg, 
but can be reached by SuperCDMS phase~C and LZ20T.  Detection of this 
$3~{\rm TeV}$ wino is possible at future detectors but the dark matter 
mass can not be determined to better than $50\%$.  The best measurement 
of the dark matter mass probably comes from indirect detection, 
especially that of monochromatic gamma lines.

The indirect detection of wino dark matter is possible due to a large 
Sommerfeld enhancement associated with the highly-degenerate and large 
wino masses.  The most promising channel is the monochromatic gamma lines, 
which we will focus on.  The cross section for $\chi^0\chi^0 \rightarrow 
\gamma\gamma$ is calculated in Ref.~\cite{Hisano:2003ec}.  For the $M_2$ 
mass range given above, the effective cross section is
\begin{equation}
  \sigma_{\chi^0\chi^0 \rightarrow \gamma\gamma} v 
  \sim (2~\mbox{--}~0.6) \times 10^{-25}~{\rm cm^3/s}.
\label{eq:wino-ID}
\end{equation}
Compared with the cross section without the Sommerfeld effect, this 
enhancement corresponds to a boost factor of $125$ ($40$) for $M_2 = 2.7$ 
($3.0$) TeV.  As shown in Figure~\ref{fig:photon-scan}, this is within the 
reach of MAGIC even for the dark matter profile at the Galactic Center much 
shallower than NFW.  The energy resolution is expected to be similar to 
lower energies, so that the mass of wino dark matter can be measured to 
$10~\mbox{--}~20\%$ in the near future.

\section{Discussion and Conclusions}
\label{sec:concl}

A description of nature that contains unexplained fine-tunings is deeply 
unsatisfactory.  There is no natural explanation for the observed size 
of the cosmological constant, and preliminary investigations of the weak 
scale have failed to uncover any sign of a natural explanation for the 
size of electroweak symmetry breaking.  Yet both of these fine-tunings 
can be understood from environmental selection on a multiverse.  The 
cosmological constant is close to a catastrophic boundary at which 
large scale structure fails to form~\cite{Weinberg:1987dv}, and the weak 
scale is close to a catastrophic boundary at which complex nuclei are 
unstable~\cite{Agrawal:1997gf}.  If the LHC fails to uncover a natural 
theory for the weak scale, then the evidence for the multiverse will be 
strengthened; if evidence is found that supports an elementary Higgs boson 
to very high energies, this strengthening will be highly significant.

An environmental weak scale is a profound change in our way of thinking 
about physics beyond the SM, and immediately leads to further questions. 
What is the scale of supersymmetry breaking, now that it has been logically 
disconnected from the weak scale?  What is to be made of the success 
of precision gauge coupling unification that arose with weak scale 
supersymmetry?  What is the nature of dark matter---in particular how 
is the WIMP hypothesis affected?  More generally, what new physics is 
expected to be within reach of the LHC and future colliders?

It is striking that the dark matter and baryon densities in our universe 
differ by only a factor of five, while their microphysical origins are 
apparently unrelated.  Furthermore, the formation of large scale structure 
is highly dependent on changes in the dark matter density relative to the 
baryon density, and it has been suggested that the dark matter density of 
our universe also arose from environmental selection~\cite{Tegmark:2005dy}. 
This leads to the view that we espouse in this paper; that the physics 
at the TeV scale depends on the nature of the dark matter that is being 
selected.  Before discussing the observable signals of the two models 
that we have studied in this paper, we give a comparison between some 
simple ways that this selection may occur in supersymmetric theories.

Suppose that at some scale $\tilde{m}$ the SM is embedded in a supersymmetric 
theory.  What are the TeV-scale effective theories that can result from 
environmental selection of both the weak scale and the dark matter abundance? 
In Table~\ref{tab:heavygauginos} (\ref{tab:lightgauginos}) we show three 
theories that have a high (low) scale of $R$ breaking in the observable 
sector.  Indeed, if the dark matter particle is either an axion or a thermal 
relic composed of states of the MSSM, these are the {\it only} such theories 
that are allowed by experiment.  (In the case of large $R$ breaking we 
also allow gauge singlet states.)
\begin{table}[t]
\begin{center}
\begin{tabular}{||c||c|c|c||}
\hline \hline
& I & II & III \\
\hline \hline
States at TeV scale & SM & (SM~$+~\tilde{w}$) & (SM~$+~\tilde{h}/\tilde{s}$) \\
\hline
Dark Matter & QCD axion & $\tilde{w}$ E-WIMP & E-WIMP LSP \\
\hline
DM selection acts on & $\theta_{\rm mis}$ & $m_{\tilde{w}}$ & $\epsilon$ \\
\hline
New parameters & $f_a, \theta_{\rm mis}$ & $m_{\tilde{w}}$ & $\mu, m, y$ \\
\hline
Gauge coupling unif. & SM & $\approx$ SM & $\approx$ MSSM \\
\hline
Higgs mass & $(128~\mbox{--}~141)~{\rm GeV}$ & 
  $(127~\mbox{--}~142)~{\rm GeV}$ & $(141~\mbox{--}~210)~{\rm GeV}$ \\
\hline \hline
\end{tabular}
\end{center}
\caption{Comparison of the three realistic TeV-scale effective theories 
 with high scale $R$ breaking that result from environmental selection 
 for dark matter in a supersymmetric theory containing the states of the 
 MSSM and possibly additional gauge singlets.  The supersymmetry breaking 
 scale is taken to be very high, broadly of order the unification scale. 
 Higgs mass values are for $m_t =173.1~{\rm GeV}$ and $\alpha_s(M_Z) 
 = 0.1176$.}
\label{tab:heavygauginos}
\end{table}
\begin{table}[t]
\begin{center}
\begin{tabular}{||c||c|c|c||}
\hline \hline
& IV & V & VI \\
\hline \hline
States at TeV scale & (SM~$+~\tilde{\lambda}$) & 
  (SM~$+~\tilde{\lambda}/\tilde{h}$) & E-MSSM \\
\hline
Dark Matter & $\tilde{w}$ E-WIMP & E-WIMP LSP & E-WIMP LSP \\
\hline
DM selection acts on & $\epsilon_R$ & $\epsilon'_R$ & $\tilde{m}$ \\
\hline
New parameters & $m_{\tilde{\lambda}}$ & 
  $m_{\tilde{\lambda}}, \mu, \tan\beta$ & MSSM set \\
\hline
Gauge coupling unif. & SM & $\approx$ MSSM & $\approx$ MSSM \\
\hline
Higgs mass & $(114~\mbox{--}~139)~{\rm GeV}$ & 
  $(114~\mbox{--}~154)~{\rm GeV}$ & $\approx (114~\mbox{--}~130)~{\rm GeV}$ \\
\hline \hline
\end{tabular}
\end{center}
\caption{Comparison of the three TeV-scale effective theories with low 
 scale $R$ breaking that result from environmental selection for dark 
 matter in a supersymmetric theory containing the states of the MSSM. 
 In all three theories the gauginos, $\tilde{\lambda}$, are at the TeV 
 scale.  In Theories~IV and V the scale of supersymmetry breaking ranges 
 from unified scales to not far above the weak scale, when these theories 
 merge with Theory~VI.  In addition to the parameters listed, Theory~V 
 (Split Supersymmetry) may involve two new physical phases.  The lower 
 bounds on the Higgs mass, $114~{\rm GeV}$, are due to experimental 
 constraints~\cite{Barate:2003sz}.}
\label{tab:lightgauginos}
\end{table}

In Theory~I dark matter is composed of axions, with environmental selection 
acting on the axion misalignment angle $\theta_{\rm mis}$~\cite{Linde:1987bx}. 
In Theories~II~--~VI, dark matter is an Environmentally selected thermal 
relic, or E-WIMP, and is the lightest supersymmetric particle (LSP), with 
selection acting on different parameters to yield the dark matter abundance. 
In Theory~II the selection acts on the mass of a particular superpartner, 
in particular the wino mass, so that there are cancellations between 
several large contributions leading to a TeV scale $\tilde{w}$, with 
all other superpartner masses of order $\tilde{m}$.  In Theory~III, the 
selection acts on the small symmetry breaking parameter $\epsilon$ of 
a non-$R$ symmetry, allowing more than one mass parameter to be small. 
The Higgs mass predictions shown in Table~\ref{tab:heavygauginos} for 
Theories~I, II and III follow from taking $\tilde{m}$ very large, within 
one or two orders of magnitude of the scale of gauge coupling unification. 
The range in the Higgs mass arises from $\tan\beta$ in Theories~I 
and II and from $y$ in Theory~III.  If the theory at scale $\tilde{m}$ 
has an approximate Peccei-Quinn symmetry, or if it is embedded 
into a higher dimensional theory with the Higgs doublets arising 
from a single supermultiplet, then the Higgs mass is predicted to 
lie at the upper end of the $\tan\beta$ range, leading to $M_H \simeq 
(141~\mbox{--}~142)~{\rm GeV}$ for Theories~I and II (and Theory~III 
with small $y$)~\cite{Hall:2009nd}.

Theories IV and V result from selection acting on an approximate $R$ 
symmetry in the observable sector, arising from a special form of the 
K\"{a}hler potential that suppresses the gaugino masses even in the 
presence of the large $U(1)_R$ breaking needed for a vanishing cosmological 
constant.  Theory~V has the $\mu$ parameter charged under this $R$ 
symmetry so that all the fermionic superpartners are at the TeV scale, 
yielding Split Supersymmetry~\cite{ArkaniHamed:2004fb}.  On the other 
hand, in Theory~IV the $\mu$ parameter is not charged under the $R$ 
symmetry, so that it is unsuppressed.  This leaves only the gauginos 
at the TeV scale.  Like Split Supersymmetry, this theory has a long-lived 
gluino, but the dark matter is necessarily pure $\tilde{w}$.  Gauge 
coupling unification depends on $\tilde{m}$, and the precision is 
typically worse than in Split Supersymmetry.  To avoid a cosmologically 
stable gluino, Theories~IV and V require $\tilde{m} \simlt 10^{10}~{\rm 
GeV}$~\cite{Arvanitaki:2005fa} or some additional light state to which 
the gluino can decay, such as an axino~\cite{Graham:2009gr}.  Hence the 
Higgs mass values quoted in Table~\ref{tab:lightgauginos} correspond 
to $\tilde{m}$ varying from the TeV scale up to the unified scale. 
Finally, Theory~VI results when the environmental selection acts on 
the supersymmetry breaking scale itself, so that all the superpartners 
are at the TeV scale.

Theories I, II and IV have gauge coupling unification with a precision 
comparable to that of the SM, while for Theories~III, V and VI the 
precision is improved and is comparable to that of the MSSM.  The 
unification scale is around $10^{14}~{\rm GeV}$ for Theories~I, II 
and III, while it is around $10^{16}~{\rm GeV}$ for Theories~IV, V 
and VI.  For Theories~II through VI, sufficient stability of the dark 
matter requires some symmetry, such as $R$ parity.

Theory~I has been studied in~\cite{Hall:2009nd}, which finds the theoretical 
uncertainties in the Higgs mass prediction to be remarkably small.  In this 
paper we have studied the Higgs mass, hadron collider and astrophysical 
signals of Theories~II and III.  The phenomenology of Split Supersymmetry, 
Theory V, has been studied in detail over a wide range of values of 
$\tilde{m}$.  The phenomenology of Theory~VI is significantly different 
than that expected for the MSSM.  In the MSSM, weak-scale supersymmetry 
is motivated to avoid a finely-tuned weak scale, and this gives constraints 
on the superpartner spectrum.  In Theory~VI the constraints on the spectrum 
are much less severe, since the only requirement is that the LSP is 
Environmentally selected dark matter; we call Theory~VI the E-MSSM. 
In the E-MSSM a little supersymmetric hierarchy is {\it predicted} 
since $\tilde{m}$ prefers to be as large as possible consistent with 
not over-producing dark matter.  For example, the dark matter could 
be dominantly Higgsino with a mass near $1~{\rm TeV}$, or dominantly 
wino with a mass near $3~{\rm TeV}$, with other superpartners of order 
$10~{\rm TeV}$.  Thus the dark matter signals of this theory may be 
similar to those of Theory~II or III.  Furthermore, all five theories 
with E-WIMP dark matter may be impossible or very difficult to probe 
at the LHC, other than via the value of the Higgs mass.

The Higgs mass is predicted very precisely in Theories~I and II for large 
$\tan\beta$, and also in Theory~III with $y \simlt 0.4$.  The central value 
is $141.0~{\rm GeV}$ in Theory~I, and just $0.54~{\rm GeV}$ ($0.35~{\rm GeV}$) 
higher in Theory~II (III).  In all three cases, the theoretical uncertainties 
are at the level of $0.4~{\rm GeV}$, while the experimental uncertainty 
of the top quark mass (QCD coupling) leads to an uncertainty in the Higgs 
mass of $\pm 1.8~{\rm GeV}$ ($\mp 1.0~{\rm GeV}$).  In Theory~III with 
$y \simgt 0.4$ the additional Yukawa interaction leads to a very significant 
increase in the Higgs mass, as shown in Figure~\ref{fig:M_H-1}.  In 
Theories~IV and V the Higgs mass is reduced by having a lower value 
of $\tilde{m}$ and additional contributions to the gauge beta functions. 
The main difference between the Higgs mass prediction in Theories~IV and 
V is that in Split Supersymmetry the Higgs boson has additional Yukawa 
couplings, increasing the Higgs mass; however, the amount of increase is 
limited because these couplings are determined by the electroweak gauge 
couplings and so not large.  A low value of the Higgs mass is expected 
in Theory~VI, although the tension with the experimental bound is not so 
tight as in the MSSM since the superpartner spectrum is less constrained. 
For example, an upper value of $\approx 125~{\rm GeV}$ ($130~{\rm GeV}$) 
occurs for top squarks of $10~{\rm TeV}$ with small (large) $A_t$.

Which observations can discriminate between these six theories, and 
convince us that one of them is indeed correct?  For Theories~I through V, 
can we get evidence for a very large amount of fine-tuning associated 
with a very high scale of supersymmetry breaking?  Here we will content 
ourselves with a comparison of the signals of Theories~I, II and III, 
which all have $\tilde{m}$ of order the unification scale $M_u \sim 
10^{14}~{\rm GeV}$.  The signals are summarized in Table \ref{tab:signals}.
\begin{table}[t]
\begin{center}
\begin{tabular}{||l||c|c|c||}
\hline \hline
& I & II & III \\
\hline \hline
States at TeV scale & SM & (SM~$+~\tilde{w}$) & 
  (SM~$+~\tilde{h}/\tilde{s}$) \\
\hline
Dark Matter & QCD axion & $\tilde{w}$ & $\tilde{h}/\tilde{s}$ \\
\hline
New parameters & $f_a, \theta_{\rm mis}$ & 
  $m_{\tilde{w}} \simeq 3~{\rm TeV}$ & $\mu, m, y$ \\
\hline
Higgs boson mass & $(128~\mbox{--}~141)~{\rm GeV}$ & 
  $(127~\mbox{--}~142)~{\rm GeV}$ & $(141~\mbox{--}~210)~{\rm GeV}$ \\
\hline \hline
DM: & no & $\sigma v \sim 10^{-25}~{\rm cm}^3 {\rm s}^{-1}$ & no \\
large indirect $\gamma$ signal && $E_\gamma \simeq 3~{\rm TeV}$ & \\
\hline
DM: & no & $\sigma_{\rm SI}^p \sim 2 \times 10^{-9}~{\rm pb}$ & 
  $\sigma_{\rm SI}^p \sim (10^{-9}~\mbox{--}~10^{-7})~{\rm pb}$ \\
direct detection signal && $m_{\rm DM} \simeq 3~{\rm TeV}$ & 
  for $100~{\rm GeV} \simlt m_{\rm DM} \simlt 800~{\rm GeV}$ \\
\hline
DM: & no & --- & 
  $\sigma v \sim (10^{-30}~\mbox{--}~10^{-28})~{\rm cm}^3 {\rm s}^{-1}$ \\
small indirect $\gamma$ signal &&& 
  except $m_{\rm D} \simlt 100~{\rm GeV}$ and \\
&&& $m_{\rm DM} \sim (200~\mbox{--}500)~{\rm GeV}$ for large $y$ \\
\hline \hline
LHC signals of new states & no & no & difficult; possible for low masses \\
\hline
signals at future lepton & no & only with & 
  precise measurements of $\mu, m, y$ \\
colliders && $\sqrt{s} \simgt 6~{\rm TeV}$ & \\
\hline \hline
\end{tabular}
\end{center}
\caption{A comparison of the signals for Theories~I, II and III.  The 
 supersymmetry breaking scale $\tilde{m}$ is of order the unification 
 scale $M_u \sim 10^{14}~{\rm GeV}$.  Beneath this scale the effective 
 theory is the SM, together with the QCD axion, $\tilde{w}$ or 
 $\tilde{h}/ \tilde{s}$, respectively.  For Theory~III the dark matter 
 detection entries are a very brief summary of Figures~\ref{fig:DM-relic}, 
 \ref{fig:dd-scan} and \ref{fig:photon-scan}.  The last three signals 
 are for the far future.}
\label{tab:signals}
\end{table}

The precision measurement of the Higgs mass at LHC is particularly important, 
since the theoretical prediction of $\simeq (141~\mbox{--}~142)~{\rm GeV}$ 
is only evaded in Theory~III with large $y$, or in Theories~I and II 
if $\tan\beta$ is relevant.  A precise measurement near $141~{\rm GeV}$ 
would by itself provide significant evidence for a very high supersymmetry 
breaking scale, no matter what the underlying theory.  If the Higgs mass 
lies above this special number by more than the uncertainties arising 
from the experimental errors in the top quark mass and QCD coupling, 
Theories~I and II will both be excluded.  A Higgs mass in the window 
$\simeq (127~\mbox{--}~142)~{\rm GeV}$ would be suggestive of Theory~I 
or II, while a value above $141~{\rm GeV}$, but below $210~{\rm GeV}$, 
would accurately determine $y$ in Theory~III.

The indirect detection of dark matter from annihilations to $\gamma\gamma$ 
at the galactic center will provide a crucial measurement.  A large 
signal at a planned atmospheric Cerenkov telescope corresponding to 
$\sigma v \sim 10^{-25}~{\rm cm}^3 {\rm s}^{-1}$ and $E_\gamma \simeq 
3~{\rm TeV}$, when combined with a Higgs mass measurement in the window 
$(127~\mbox{--}~142)~{\rm GeV}$, would provide strong evidence for 
Theory~II.  This would be compelling for a Higgs mass at the upper 
end of the window, implying $\sim 24$~orders of magnitude fine-tuning. 
For Theory~III, Figure~\ref{fig:photon-scan} shows that, for a wide 
range of dark matter masses, $100~{\rm GeV} \simlt m_{\rm DM} \simlt 
1~{\rm TeV}$, a monochromatic photon signal from the galactic center 
should be seen with $\sigma v \simgt 10^{-30}~{\rm cm}^3 {\rm s}^{-1}$. 
Except for very small regions of parameter space at low $m_{\rm DM}$ 
and low $y$, a smaller value of $\sigma v$ only occurs at large $y$, 
and this must be correlated with a heavy Higgs.  Thus the combination 
of the Higgs mass and monochromatic photon measurements serves to 
essentially distinguish the three theories.

Further discrimination between the theories is provided by direct 
detection of dark matter.  Figure~\ref{fig:dd-scan} shows that in 
Theory~III with $m_{\rm DM} < 800~{\rm GeV}$ a signal with $\sigma v 
\sim (10^{-9}~\mbox{--}~10^{-7})~{\rm pb}$ is expected for any value 
of $y$.  While none of the signal region is presently excluded, much 
is within reach of XENON~100kg and LUX~300kg.  Theory~II predicts 
$\sigma_{\rm SI}^p \simeq 2 \times 10^{-9}~{\rm pb}$ and, since 
$m_{\rm DM} \simeq 3~{\rm TeV}$, experiments beyond XENON~100kg 
and LUX~300kg will be needed.

In Theories I and II there is no possibility of any signal of new physics 
beyond the SM Higgs at the LHC.  A trilepton signal for Theory~III is 
possible, and the most favorable region of parameter space is low $\mu$ 
and $m$ and also low $y$.  However, with an integrated luminosity of 
$100~{\rm fb}^{-1}$, the number of signal events is at most of order 
$100$, and the background is about a factor $3$ larger, so that even 
in this favorable case it is not clear that $y$ can be extracted 
sufficiently accurately to allow a precise correlation with the Higgs 
mass.  For Theory~III a precise determination of parameters via the 
Higgsino/singlino masses and couplings would require a linear $e^+e^-$ 
collider.  If the dark matter mass is approaching $1~{\rm TeV}$, a muon 
collider may be more effective.

While a Higgs mass near $141~{\rm GeV}$ would point towards Theory~I, 
it would become convincing only after excluding Theories~II and III. 
The absence of an indirect detection $\gamma$ signal with $\sigma v 
> 10^{-26}~{\rm cm}^3 {\rm s}^{-1}$ at $3~{\rm TeV}$ would clearly 
exclude Theory~II.  For Theory~III, the absence of a direct detection 
signal with $\sigma v > 10^{-9}~{\rm pb}$ would essentially exclude 
$100~{\rm GeV} \simlt m_{\rm DM} \simlt 800~{\rm GeV}$.  Theory~III with 
$m_{\rm DM} \simgt 800~{\rm GeV}$, the case of dominant Higgsino dark 
matter, would be excluded by the absence of an indirect $\gamma$ signal 
with $\sigma v > 10^{-29}~{\rm cm}^3 {\rm s}^{-1}$, while that with 
$m_{\rm DM} \simlt 100~{\rm GeV}$ may be within the reach of the LHC.

The main theoretical motivation for the LHC is to uncover the physics 
behind the weak scale by discovering new particles beyond the SM. 
However, this motivation assumes that the weak scale is naturally 
related to some other mass scale, such as the scale of some new strong 
force or of supersymmetry breaking, without any fine-tuning.  If the 
weak scale arises from environmental selection on a multiverse, the 
most significant evidence from the LHC may lie in the value of the 
Higgs boson mass.  Combining this with results from the direct and 
indirect detection of galactic dark matter could convincingly uncover 
the nature of the weak scale theory, providing strong evidence for 
both the multiverse and supersymmetry at unified scales.

\section*{Acknowledgments}

P.K. thanks Hitoshi Murayama and Tomer Volansky for useful discussions. 
This work was supported in part by the Director, Office of Science, 
Office of High Energy and Nuclear Physics, of the US Department of 
Energy under Contract DE-AC02-05CH11231, and in part by the National 
Science Foundation under grants PHY-0457315, PHY-0555661 and PHY-0855653.


\begin{thebibliography}{99}

\bibitem{Perlmutter:1998np}
S.~Perlmutter {\it et al.}  [Supernova Cosmology Project Collaboration],
Astrophys.\ J.\  {\bf 517}, 565 (1999)
[arXiv:astro-ph/9812133];
A.~G.~Riess {\it et al.}  [Supernova Search Team Collaboration],
Astron.\ J.\  {\bf 116}, 1009 (1998)
[arXiv:astro-ph/9805201].

\bibitem{Komatsu:2008hk}
E.~Komatsu {\it et al.}  [WMAP Collaboration],
Astrophys.\ J.\ Suppl.\  {\bf 180}, 330 (2009)
[arXiv:0803.0547 [astro-ph]];
M.~Kowalski {\it et al.}  [Supernova Cosmology Project Collaboration],
Astrophys.\ J.\  {\bf 686}, 749 (2008)
[arXiv:0804.4142 [astro-ph]].

\bibitem{Weinberg:1987dv}
S.~Weinberg,
Phys.\ Rev.\ Lett.\  {\bf 59}, 2607 (1987).

\bibitem{Agrawal:1997gf}
V.~Agrawal, S.~M.~Barr, J.~F.~Donoghue and D.~Seckel,
Phys.\ Rev.\  D {\bf 57}, 5480 (1998)
[arXiv:hep-ph/9707380];
T.~Damour and J.~F.~Donoghue,
Phys.\ Rev.\  D {\bf 78}, 014014 (2008)
[arXiv:0712.2968 [hep-ph]].

\bibitem{Bousso:2000xa}
R.~Bousso and J.~Polchinski,
JHEP {\bf 06}, 006 (2000)
[arXiv:hep-th/0004134];
S.~Kachru, R.~Kallosh, A.~Linde and S.~P.~Trivedi,
Phys.\ Rev.\  D {\bf 68}, 046005 (2003)
[arXiv:hep-th/0301240];
L.~Susskind,
arXiv:hep-th/0302219;
M.~R.~Douglas,
JHEP {\bf 05}, 046 (2003)
[arXiv:hep-th/0303194].

\bibitem{Tegmark:2005dy}
For recent discussions on environmental selection for dark matter, see, e.g.,
M.~Tegmark, A.~Aguirre, M.~J.~Rees and F.~Wilczek,
Phys.\ Rev.\  D {\bf 73}, 023505 (2006)
[arXiv:astro-ph/0511774];
S.~Hellerman and J.~Walcher,
Phys.\ Rev.\  D {\bf 72}, 123520 (2005)
[arXiv:hep-th/0508161].

\bibitem{ArkaniHamed:2004fb}
N.~Arkani-Hamed and S.~Dimopoulos,
JHEP {\bf 0506}, 073 (2005)
[arXiv:hep-th/0405159];
G.~F.~Giudice and A.~Romanino,
Nucl.\ Phys.\  B {\bf 699}, 65 (2004)
[Erratum-ibid.\  B {\bf 706}, 65 (2005)]
[arXiv:hep-ph/0406088];
N.~Arkani-Hamed, S.~Dimopoulos, G.~F.~Giudice and A.~Romanino,
Nucl.\ Phys.\  B {\bf 709}, 3 (2005)
[arXiv:hep-ph/0409232].

\bibitem{ArkaniHamed:2005yv}
N.~Arkani-Hamed, S.~Dimopoulos and S.~Kachru,
arXiv:hep-th/0501082.

\bibitem{Mahbubani:2005pt}
R.~Mahbubani and L.~Senatore,
Phys.\ Rev.\  D {\bf 73}, 043510 (2006)
[arXiv:hep-ph/0510064].

\bibitem{Hall:2001pg}
L.~J.~Hall and Y.~Nomura,
Phys.\ Rev.\  D {\bf 64}, 055003 (2001)
[arXiv:hep-ph/0103125].

\bibitem{Hall:2001xb}
L.~J.~Hall and Y.~Nomura,
Phys.\ Rev.\  D {\bf 65}, 125012 (2002)
[arXiv:hep-ph/0111068].

\bibitem{Hall:2009nd}
L.~J.~Hall and Y.~Nomura,
arXiv:0910.2235 [hep-ph].

\bibitem{Tevatron:2009ec}
The Tevatron Electroweak Working Group,
arXiv:0903.2503 [hep-ex].

\bibitem{Amsler:2008zzb}
C.~Amsler {\it et al.}  [Particle Data Group],
Phys.\ Lett.\  B {\bf 667}, 1 (2008),
and 2009 partial update for the 2010 edition.

\bibitem{Belanger:2007zz}
G.~Belanger, F.~Boudjema, A.~Pukhov and A.~Semenov,
Comput.\ Phys.\ Commun.\  {\bf 177}, 894 (2007).

\bibitem{Komatsu:2008hk-2}
E.~Komatsu {\it et al.}  [WMAP Collaboration],
in Ref.~\cite{Komatsu:2008hk}.

\bibitem{Hisano:2006nn}
J.~Hisano, S.~Matsumoto, M.~Nagai, O.~Saito and M.~Senami,
Phys.\ Lett.\  B {\bf 646}, 34 (2007)
[arXiv:hep-ph/0610249].

\bibitem{CDFNote}
CDFNote-9713

\bibitem{Collaboration:2009je}
The TEVNPH Collaboration,
arXiv:0911.3930 [hep-ex].

\bibitem{Abdullin:2005yn}
S.~Abdullin {\it et al.},
Eur.\ Phys.\ J.\  C {\bf 39S2}, 41 (2005).

\bibitem{Ball:2007zza}
G.~L.~Bayatian {\it et al.}  [CMS Collaboration],
J.\ Phys.\ G {\bf 34}, 995 (2007).

\bibitem{ATLAS:1999fr}
ATLAS detector and physics performance. Technical design report. Vol.~2.

\bibitem{Enberg:2007rp}
R.~Enberg, P.~J.~Fox, L.~J.~Hall, A.~Y.~Papaioannou and M.~Papucci,
JHEP {\bf 0711}, 014 (2007)
[arXiv:0706.0918 [hep-ph]].

\bibitem{Aad:2009wy}
G.~Aad {\it et al.}  [The ATLAS Collaboration],
arXiv:0901.0512 [hep-ex].

\bibitem{Vandelli:2007zz}
W.~Vandelli, CERN-THESIS-2007-072.

\bibitem{Hisano:2004pv}
J.~Hisano, S.~Matsumoto, M.~M.~Nojiri and O.~Saito,
Phys.\ Rev.\  D {\bf 71}, 015007 (2005)
[arXiv:hep-ph/0407168].

\bibitem{DM-tool}
R.~Gaitskell and J.~Filippini, {\tt http://dendera.berkeley.edu/plotter/entryform.html},
and references therein.

\bibitem{Shan:2009ym}
C.-L.~Shan,
New J.\ Phys.\  {\bf 11}, 105013 (2009)
[arXiv:0903.4320 [hep-ph]].

\bibitem{Drees:2008bv}
M.~Drees and C.-L.~Shan,
JCAP {\bf 0806}, 012 (2008)
[arXiv:0803.4477 [hep-ph]].

\bibitem{Green:2008rd}
A.~M.~Green,
JCAP {\bf 0807}, 005 (2008)
[arXiv:0805.1704 [hep-ph]].

\bibitem{Bergstrom:1997fj}
L.~Bergstr\"{o}m, P.~Ullio and J.~H.~Buckley,
Astropart.\ Phys.\  {\bf 9}, 137 (1998)
[arXiv:astro-ph/9712318].

\bibitem{Hisano:2003ec}
J.~Hisano, S.~Matsumoto and M.~M.~Nojiri,
Phys.\ Rev.\ Lett.\  {\bf 92}, 031303 (2004)
[arXiv:hep-ph/0307216];
J.~Hisano, S.~Matsumoto, M.~M.~Nojiri and O.~Saito,
Phys.\ Rev.\  D {\bf 71}, 063528 (2005)
[arXiv:hep-ph/0412403].

\bibitem{Buckley:2008ud}
J.~Buckley {\it et al.},
arXiv:0810.0444 [astro-ph];
{\tt http://www.cta-observatory.org}

\bibitem{Arvanitaki:2004df}
A.~Arvanitaki and P.~W.~Graham,
Phys.\ Rev.\  D {\bf 72}, 055010 (2005)
[arXiv:hep-ph/0411376].

\bibitem{Baltz:2008wd}
E.~A.~Baltz {\it et al.},
JCAP {\bf 0807}, 013 (2008)
[arXiv:0806.2911 [astro-ph]].

\bibitem{Aharonian:2006wh}
F.~Aharonian {\it et al.}  [H.E.S.S. Collaboration],
Phys.\ Rev.\ Lett.\  {\bf 97}, 221102 (2006)
[Erratum-ibid.\  {\bf 97}, 249901 (2006)]
[arXiv:astro-ph/0610509].

\bibitem{Serpico:2008ga}
P.~D.~Serpico and G.~Zaharijas,
Astropart.\ Phys.\  {\bf 29}, 380 (2008)
[arXiv:0802.3245 [astro-ph]].

\bibitem{Bertone:2005hw}
G.~Bertone and D.~Merritt,
Phys.\ Rev.\  D {\bf 72}, 103502 (2005)
[arXiv:astro-ph/0501555].

\bibitem{Bertone:2009cb}
G.~Bertone, C.~B.~Jackson, G.~Shaughnessy, T.~M.~P.~Tait and A.~Vallinotto,
Phys.\ Rev.\  D {\bf 80}, 023512 (2009)
[arXiv:0904.1442 [astro-ph.HE]].

\bibitem{Pierce:1996zz}
D.~M.~Pierce, J.~A.~Bagger, K.~T.~Matchev and R.-J.~Zhang,
Nucl.\ Phys.\  B {\bf 491}, 3 (1997)
[arXiv:hep-ph/9606211];
H.-C.~Cheng, B.~A.~Dobrescu and K.~T.~Matchev,
Nucl.\ Phys.\  B {\bf 543}, 47 (1999)
[arXiv:hep-ph/9811316].

\bibitem{Essig:2007az}
R.~Essig,
Phys.\ Rev.\  D {\bf 78}, 015004 (2008)
[arXiv:0710.1668 [hep-ph]];
M.~Cirelli and A.~Strumia,
New J.\ Phys.\  {\bf 11}, 105005 (2009)
[arXiv:0903.3381 [hep-ph]].

\bibitem{Barate:2003sz}
R.~Barate {\it et al.}  [ALEPH Collaboration],
Phys.\ Lett.\ B {\bf 565}, 61 (2003)
[arXiv:hep-ex/0306033];
LEP Higgs Working Group Collaboration,
arXiv:hep-ex/0107030.

\bibitem{Linde:1987bx}
A.~D.~Linde,
Phys.\ Lett.\  B {\bf 201}, 437 (1988);
F.~Wilczek,
arXiv:hep-ph/0408167;
M.~Tegmark, A.~Aguirre, M.~J.~Rees and F.~Wilczek,
in Ref.~\cite{Tegmark:2005dy}.

\bibitem{Arvanitaki:2005fa}
A.~Arvanitaki, C.~Davis, P.~W.~Graham, A.~Pierce and J.~G.~Wacker,
Phys.\ Rev.\  D {\bf 72}, 075011 (2005)
[arXiv:hep-ph/0504210].

\bibitem{Graham:2009gr}
P.~W.~Graham and S.~Rajendran,
arXiv:0906.4657 [hep-ph].

\end{thebibliography}
\end{document}